\documentclass{aanoln}

\usepackage{verbatim}
\usepackage{graphicx}
\usepackage{xcolor}
\usepackage{booktabs}
\usepackage[colorlinks=true,linkcolor=blue,urlcolor=blue,citecolor=blue]{hyperref}
\usepackage{xspace}
\usepackage{pdflscape}
\usepackage{subcaption}
\usepackage{placeins}
\usepackage{longtable}
\usepackage{threeparttablex}

\usepackage{txfonts}\usepackage{twoopt}
\bibpunct{(}{)}{;}{a}{}{,}
\usepackage{enumitem}
\defcitealias{schneider2024a}{Paper~I}
\defcitealias{hamuy2003b}{H03}
\defcitealias{martinez2022a}{M22}
\defcitealias{utrobin2024a}{UC24}

\graphicspath{{./}, {figures/}}

\newcommand{\msun}{{\rm M}_{\odot}}
\newcommand{\lsun}{{\rm L}_{\odot}}
\newcommand{\rsun}{{\rm R}_{\odot}}

\newcommand{\kms}{{\rm km\,s^{-1}}}

\newcommand{\mesa}{\mbox{\textsc{Mesa}}\xspace}
\newcommand{\eg}{e.g.\@\xspace}
\newcommand{\cf}{c.f.\@\xspace}
\newcommand{\ie}{i.e.\@\xspace}
\newcommand{\snesa}{SN~1987A\xspace}
\newcommand{\snesalike}{SN~1987A-like\xspace}
\newcommand{\snesalikeshort}{87A-like\xspace}
\newcommand{\snii}{SN~II\xspace}
\newcommand{\sniip}{SN~IIP\xspace}
\newcommand{\sniin}{SN~IIn\xspace}
\newcommand{\sniil}{SN~IIL\xspace}
\newcommand{\sniib}{SN~IIb\xspace}
\newcommand{\snibc}{SN~Ib/c\xspace}

\newcommand{\sneii}{SNe~II\xspace}
\newcommand{\sneiip}{SNe~IIP\xspace}
\newcommand{\sneiin}{SNe~IIn\xspace}
\newcommand{\sneiil}{SNe~IIL\xspace}
\newcommand{\sneiib}{SNe~IIb\xspace}
\newcommand{\sneibc}{SNe~Ib/c\xspace}

\newcommand{\casea}{Case~A\@\xspace}
\newcommand{\caseae}{Case~A-e\@\xspace}
\newcommand{\caseal}{Case~A-l\@\xspace}
\newcommand{\caseb}{Case~B\@\xspace}
\newcommand{\casebe}{Case~B-e\@\xspace}
\newcommand{\casebl}{Case~B-l\@\xspace}
\newcommand{\casec}{Case~C\@\xspace}

\newcommand{\caseah}{Case-A\@\xspace}

\newcommand{\casebh}{Case-B\@\xspace}

\newcommand{\casech}{Case-C\@\xspace}

\titlerunning{Supernovae from stellar mergers and accretors of binary mass transfer}
\authorrunning{F.R.N.~Schneider et al.}
\begin{document}
\title{Supernovae from stellar mergers and accretors of binary mass transfer: Implications for Type~IIP, 1987A-like and interacting supernovae}
\author{%
    F.R.N.~Schneider\inst{\ref{HITS},\ref{ZAH}}\thanks{fabian.schneider@h-its.org}
    \and E.~Laplace\inst{\ref{HITS},\ref{KUL},\ref{AMS}}
    \and Ph.~Podsiadlowski\inst{\ref{LISA},\ref{HITS},\ref{OXFORD}}
}
\institute{%
    Heidelberger Institut f{\"u}r Theoretische Studien, Schloss-Wolfsbrunnenweg 35, 69118 Heidelberg, Germany\label{HITS}
    \and Zentrum f{\"u}r Astronomie der Universit{\"a}t Heidelberg, Astronomisches Rechen-Institut, M{\"o}nchhofstr.\ 12-14, 69120 Heidelberg, Germany\label{ZAH}
    \and Institute of Astronomy, KU Leuven, Celestijnenlaan 200D, 3001 Leuven, Belgium\label{KUL}
    \and Anton Pannekoek Institute of Astronomy, Science Park 904, University of Amsterdam, 1098XH Amsterdam, The Netherlands\label{AMS}
    \and London Institute of Stellar Astrophysics, Vauxhall, London, United Kingdom\label{LISA}
    \and University of Oxford, St Edmund Hall, Oxford, OX1 4AR, United Kingdom\label{OXFORD}
}
\date{Received 04 March 2025 / Accepted 30 Juni 2025}
\abstract{As most massive stars are born in binary and other multiple-star systems, many are expected to exchange mass with a companion star or merge with it during their lives. This means that most supernovae (SNe) are from such binary products. Here, we focus on hydrogen-rich Type~II SNe from accretors of binary mass transfer and stellar mergers, and contrast them to those from single stars. We compute various SN properties such as the explosion energies, nickel yields, and neutron star (NS) kick velocities, but also consider NS masses. We find tight correlations between these parameters and various summary variables of the pre-SN core structures of stars such as the central specific entropy, core compactness, and iron core mass. However, there is no obvious relation between these explosion properties and the evolutionary history of the pre-SN stars (\ie single stars vs.\ binary mass accretors and stellar mergers). We find linear relations between the nickel mass and the SN explosion energy and the NS remnant mass, and motivate the reasons for such relations in our models. These relations allow, in principle, for the determination of SN explosion energies and NS masses from nickel masses, \eg, inferred from the tail of SN light curves. We further group our models into progenitors of \sneiip, \snesalike and interacting SNe, predict their SN and SN-progenitor properties and compare these to observations. Overall, there is good agreement, but we also highlight some tension. Accretors of binary mass transfer and stellar mergers naturally produce \sneiip with long plateau durations from progenitors with relatively small CO-cores but large envelope masses that could explain SNe such as SN~2015ba. Our models give rise to tight relations between the plateau luminosity and the nickel mass as well as the SN ejecta velocity as inferred observationally for \sneiip. We speculate that cool/red supergiants at $\log\,L/\lsun\,{\geq}\,5.5$ encounter enhanced mass loss due to envelope instabilities and that some could retain a hydrogen envelope to then explode in interacting \sneiin. The rate of such SNe from our models seems compatible with the observed \sniin rate. Some of our binary models explode as $10^6\,\lsun$ blue supergiants that may have encountered enhanced and/or eruptive mass loss shortly before their SNe and could thus help understand interacting SNe such as SN~1961V and SN~2005gl but also superluminous Type~II SNe such as SN~2010jl.}
\maketitle
%
%
%
%
\section{\label{sec:introduction}Introduction}

The vast majority of massive stars and hence supernova (SN) progenitors are born in multiple stellar systems and will exchange mass with a companion during their lifespans \citep[\eg][]{sana2012a, moe2017a}. Mass exchange can take on different forms and stars can be stripped off their outer envelopes, giving rise to hydrogen-free SNe of Types~Ib/c, but they can also gain mass by stable Roche-lobe overflow and merging \citep[see, \eg, reviews by][]{langer2012a, marchant2024a}. In the latter cases, stars typically explode in hydrogen-rich SNe of Type~II\footnote{In very massive post-mass accretors, stellar winds and other forms of mass loss may again lead to \snibc.}.

It is estimated that about half of all \snii are from progenitor stars that accreted mass or merged with a companion star, while some 15\% are thought to originate from mergers involving a main-sequence (MS) and a post-MS star \citep[\eg][]{podsiadlowski1992a, zapartas2019a, henneco2024a}. These post-MS+MS star mergers can explain the unusual appearance of the progenitor of \snesa as a blue supergiant (BSG), the peculiar SN light curve and also the triple-ring nebula \citep[\eg][]{podsiadlowski1990a, podsiadlowski1992c, podsiadlowski2017a, morris2007a, morris2009a, menon2017a, menon2019a}. Blue supergiants from stellar mergers may also explode in interacting, superluminous and pair-instability SNe or collapse into very massive black holes (BHs) (see, \eg, \citealt{justham2014a, dicarlo2020a, renzo2020a, costa2022a}; \citealt{schneider2024a}, hereafter \citetalias{schneider2024a}).

Binary-stripped stars (BSSs) and explosions thereof as \snibc are important for many areas in astrophysics and have been studied in quite some detail \citep[\eg][]{paczynski1967b, vandenheuvel1976a, deloore1978a, vanbeveren1991a, podsiadlowski1992a, wellstein1999a, petrovic2005a, cantiello2007a, eldridge2008a, yoon2010a, yoon2017a, tauris2013a, tauris2015a, langer2020a, laplace2020a, bavera2021a, schneider2021a, schneider2023a, klencki2022a, wang2024a}. For example, BSSs contributed ${\approx}\,20\%$ of the photons to the reionisation of hydrogen in the Universe after the Dark Ages \citep{gotberg2020a}, and many of the compact objects in gravitational-wave (GW) merger events from isolated binary stars are from such objects (see, \eg, schematic evolutionary cartoon in figure~1 of \citealt{vanson2022c} and the reviews by \citealt{mapelli2021a} and \citealt{mandel2022a}).

Accretors of stable mass transfer and stellar mergers are somewhat less studied but similarly important. They likely form some of the most massive stars in the Universe and become massive counterparts of classical blue stragglers. Some of the most massive stars known, the $200\text{--}300\,\msun$ stars in R136 \citep[][]{crowther2010a, bestenlehner2020a, brands2022a}, may be such objects \citep[see, \eg,][]{banerjee2012b, schneider2014a}. Moreover, accretors and mergers stem from initially lower-mass stars with long lifetimes and additionally rejuvenate such that they can explode with long delay times \citep[][]{zapartas2017a}. After the SN of their companion, accretors may become runaway stars \citep[\eg][]{blaauw1961a, renzo2019a}, and all of these effects can significantly enhance stellar feedback \citep[\eg][]{conroy2012a, kimm2014a, ma2016a, secunda2020a, geen2023a}. Furthermore, the large diversity of the light curves of hydrogen-rich Type~II SNe would be difficult to explain without mass accretion in binary stars \citep[see, \eg,][]{eldridge2018a}.

In \citetalias{schneider2024a}, we modelled accretors and stellar mergers, and established how their cores differ from those of genuine single stars right at iron core collapse. We paid special attention to their pre-SN core structures, likely fate, compact remnants, and SN ejecta masses. In particular, accretors and mergers involving post-MS stars can become long-lived BSGs \citep[\eg][]{hellings1983a,hellings1984a, podsiadlowski1989a, claeys2011a, vanbeveren2013a, justham2014a, menon2017a}, and we have shown that --- at the pre-SN stage --- their cores and total masses differ most from those of single stars. Consequently, some of these long-lived BSGs may form BHs of up to $50\,\msun$ while others eject up to $40\,\msun$ in SN explosions even at solar metallicity. In this paper, we build on the models of \citetalias{schneider2024a} and explore how different the SN properties, such as the explosion energies, nickel yields and neutron-star (NS) kick velocities, are in accretors compared to genuine single stars. We further link our models to the likely SN type and explore how accretors and mergers contribute to the observed large diversity of hydrogen-rich Type~II SNe. To this end, we assign the SN types of \sniip, \snesalike and interacting \sniin according to the pre-SN location of our models in the Hertzsprung--Russell (HR) diagram.

This paper is structured as follows. We introduce the stellar and SN explosion models in Sect.~\ref{sec:methods} before presenting our findings on the SN explosion properties and the emerging SN diversity in Sect.~\ref{sec:results}. In Sect.~\ref{sec:comparison-observations}, we compare our models to observed SNe of the three broad types of \sniip, \snesalike SNe and interacting \sniin. We summarise and conclude in Sect.~\ref{sec:conclusions}.

%
%
\section{\label{sec:methods}Methods}

The models used in this paper were computed and described in \citet{schneider2021a} and \citetalias{schneider2024a}. We give a brief overview of them here and refer to \citetalias{schneider2024a} for more details. The stellar models employed revision 10398 of the Modules-for-Experiments-in-Stellar-Astrophysics (\mesa) software package \citep[][]{paxton2011a, paxton2013a, paxton2015a, paxton2018a, paxton2019a}. In Table~\ref{tab:models}\footnote{An electronic version of this table is available at \url{https://doi.org/10.5281/zenodo.15791044}.}, we provide the newly derived quantities from the stellar models used in this study and also list some previously published quantities that will aid the understanding of this paper.

The stellar models are for single stars and accretors of binary mass transfer and stellar mergers. Both the mass accretors and mergers were approximated by accretion onto single-star models with the momentary thermal-timescale mass accretion rate. Our models had initial masses $M_\mathrm{ini}$ in the range $11\text{--}70\,\msun$, and the amount of accreted mass $\Delta M_\mathrm{acc}$ was 10\%, 25\%, 50\%, 75\%, 100\%, 150\% and 200\% of a star's initial mass (denoted by the mass accretion fraction $f_\mathrm{acc}=\Delta M_\mathrm{acc}/M_\mathrm{ini}$). The mass accretion took place at five different evolutionary phases: during core hydrogen burning (called \casea accretors here), after the main sequence but before core helium ignition (called \caseb accretors), and after core helium exhaustion (called \casec accretors). \casea and~B accretors are further divided into ``early'' and ``late''. In \casea accretors, this means accretion when the central hydrogen mass fraction dropped to 0.35 (early) and 0.05 (late). In early \caseb accretors, we accreted mass shortly after the star left the main sequence, \ie while it had a radiative envelope, and in late \caseb accretors, mass accretion only occurred on the red supergiant branch when the stars had developed a deep convective envelope. With these accretion fractions and evolutionary phases of accretion, our model grid covers a large range of possible outcomes of accretors of binary mass transfer and stellar mergers. In total, 419 single and accretor models were computed.

The models were all non-rotating and employed a solar-like chemical composition \citep[\ie initial hydrogen and helium mass fractions of $X=0.7155$ and $Y=0.2703$, respectively, and metallicity $Z=0.0142$;][]{asplund2009a}. Convective step overshooting of 0.2 pressure scale heights was applied during core hydrogen and core helium burning but not in the later nuclear-burning stages. Semi-convection was included with an efficiency factor of $\alpha_\mathrm{sc}=0.1$, and a slightly modified version of \mesa's ``Dutch'' wind mass loss was used\footnote{The modifications were only for the metallicity-scaling of winds.}. Importantly, we did not use enhanced winds for stars possibly evolving into luminous-blue variables (LBVs). A 23-isotope nuclear reaction network (\texttt{approx21\_cr60\_plus\_co56.net}) was applied. All models were computed from the zero-age main sequence until iron core collapse, \ie until the iron core reaches infall velocities ${>}\,950\,\kms$.

The outcome of iron core collapse was then studied with the parametric SN model of \citet{muller2016a} with calibrations as described in \citet{schneider2021a}. This neutrino-driven SN model takes as input the entire interior structure of stars at core collapse and does not rely only on certain summary variables such as the compactness of the core structures of SN progenitors. Moreover, it allows us to study the explodability of the models, \ie the question of neutron-star (NS) vs.\ black-hole (BH) formation, and also predicts the SN explosion energy $E_\mathrm{expl}$, the ejected nickel mass $M_\mathrm{Ni}$ and the NS kick velocity. 

As explained in \citetalias{schneider2024a}, our accretor models cannot capture all the complex physics of binary mass accretion and stellar mergers but provide a useful first approximation of the interior structure and post-accretion evolution of such stars. There are four main limitations: Firstly, accretion on a thermal timescale is not fully adequate for all binary mass transfer phases and stellar mergers. \caseah mass transfer also proceeds on a nuclear timescale during which the accretor can continue to evolve. \casebh and-C mass transfer are more adequately matched by our assumptions. Stellar mergers proceed on a dynamic timescale. We cannot replicate the complex physics of such phases, but we can model the most important aspect, namely, the increase in mass. In this respect, our models do a decent job as thermal-timescale accretion does not allow the accretor to change its chemical structure by nuclear burning. Secondly, as already mentioned, the complex physics of stellar mergers cannot be reproduced by accretion alone. In particular, there is mixing in mergers, \eg, of nuclearly-processed material from the core into the envelope and vice versa \citep[\cf multi-dimensional hydrodynamic simulations of stellar mergers by, \eg,][]{lombardi2002a, ivanova2002a, ivanova2002c, glebbeek2013a, schneider2019a}. Additional mixing of helium out of the cores of mergers enhances the chance of forming long-lived BSGs \citep[\eg][]{hillebrandt1989a, podsiadlowski1990a, podsiadlowski1992c, podsiadlowski2017a, podsiadlowski2017b, menon2017a}, and our models thus underpredict their rates. Similarly, our accretion models cannot correctly predict the surface chemical abundances. Thirdly, the overall mass loss in a merger event is uncertain and predictions thereof depend, \eg, on how stars merge \citep[head-on collisions of different collision energies vs.\ less violent inspiral mergers of circular binaries; see, \eg,][]{freitag2005a, glebbeek2013a, schneider2019a}. While these uncertainties are important they do not affect our results, because we do not model individual binary systems but only effective mass accretion events of certain masses. By construction, we always have a net mass gain, but, in late \casebh and \casech mergers, the merger dynamics may lead to a net mass loss \citep[\cf classical common-envelope evolution where the entire envelope of a (super)giant star may be ejected;][]{ivanova2013a, ropke2023a, schneider2025a}. Such situations could lead to SN progenitors with little ejecta mass (\eg \sniib) that our models do not cover. Lastly, we neglect rotation while it is thought to influence the further evolution of accretors of binary mass transfer because of induced mixing. Merger products may be slowly rotating \citep[\eg][]{leonard1995a, schneider2019a, schneider2020a}, but some could also rotate more rapidly. Moreover, strong, large-scale magnetic fields are thought to be formed in stellar mergers \citep[\eg][]{schneider2019a, frost2024a, ryu2025a} and they will further influence the evolution and rotation of merged stars via enhanced angular momentum transport and magnetic braking \citep[\eg][]{schneider2020a}. Understanding the influence of all of these processes on the pre-SN structures and core-collapse outcomes is left to future studies.

%
%
\section{\label{sec:results}Results}

\begin{figure*}
    \centering
    \includegraphics{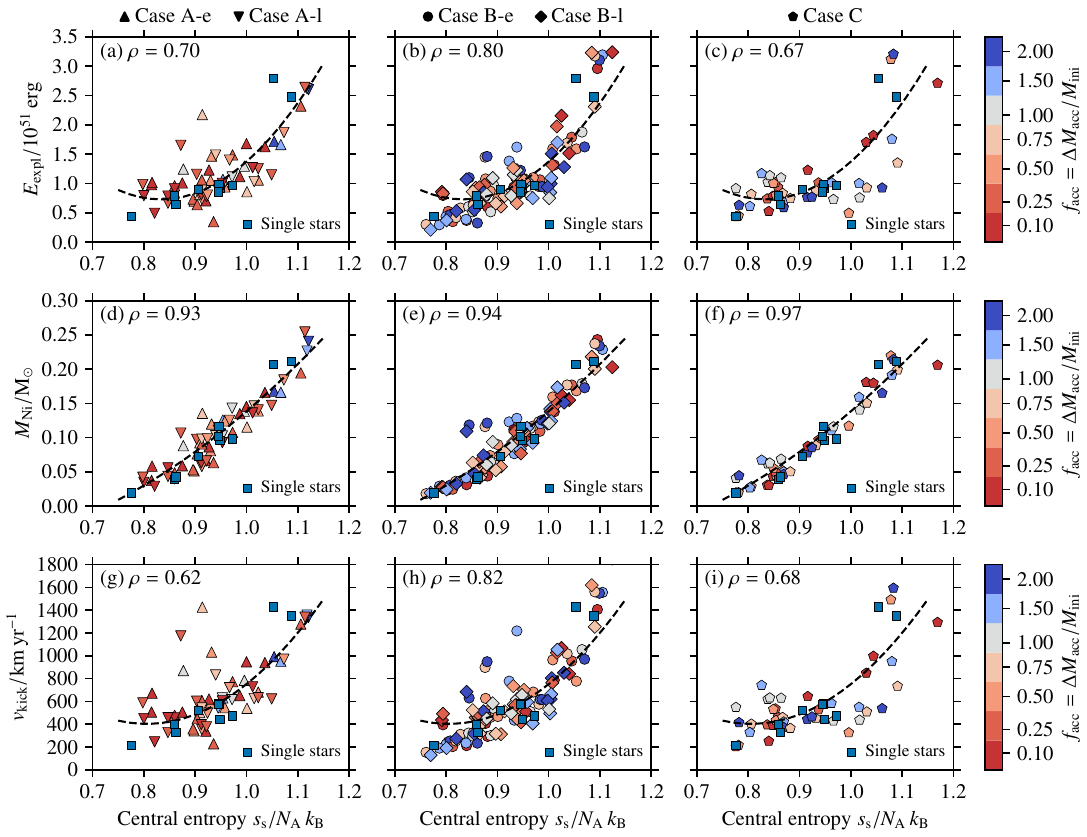}
    \caption{Explosion energy $E_\mathrm{expl}$, nickel mass $M_\mathrm{Ni}$ and NS kick velocity $v_\mathrm{kick}$ as a function of central specific entropy $s_\mathrm{c}$ of the SN progenitors of \caseah (panels a, d, g), \casebh (panels b, e, h) and \casech accretors (panels c, f, i). Colours indicate the amount of accreted mass $f_\mathrm{acc}$, and results for single-star models are shown for comparison. Pearson's correlation coefficients $\rho$ are provided, and to guide the eye and to allow for easier comparison of the data across panels, quadratic curves fitted to the data are added.}
    \label{fig:Eexpl-MNi-vkick}
\end{figure*}

We first present the predicted SN explosion outcomes in terms of the explosion energy, nickel yields and NS kick velocities (Sect.~\ref{sec:explosion-properties}) and identify the nickel mass as a proxy of the NS remnant mass and SN explosion energy (Sect.~\ref{sec:nickel-proxy}). Based on the position of our SN progenitor models in the HR diagram, we then associate them with their most likely SN types (\sniip, \snesalike and \sniin; Sect.~\ref{sec:hrd}). While this cannot be an accurate classification, it nevertheless allows us to compare the qualitative differences in the SN outcomes and properties between these three groups (Sect.~\ref{sec:sn-diversity}). Lastly, we focus on \sneiip and their expected light curve properties given the diversity of SN progenitors (Sect.~\ref{sec:light curves}).

\subsection{\label{sec:explosion-properties}Explosion energies, nickel masses and NS kicks}

We now consider the explosion energy $E_\mathrm{expl}$, nickel mass $M_\mathrm{Ni}$ and kick velocity $v_\mathrm{kick}$ imparted onto the NS obtained by applying the parametric SN code of \citet[][\cf Sect.~\ref{sec:methods}]{muller2016a} to the core-collapse structures of our single-star and accretor models. We neither find obvious correlations between the SN properties and the CO core mass $M_\mathrm{CO}$ nor the amount of accreted mass $f_\mathrm{acc}$ (\cf Table~\ref{tab:models}). 

\citet{temaj2024a} and \citetalias{schneider2024a} showed that the central specific entropy is a good proxy to summarise the core structure of stars, and, for example, they found tight correlations of central entropy with the gravitational NS mass for a large range of different SN progenitors (single stars with various degrees of convective boundary mixing and binary accretors \& stellar mergers). Here, we use this proxy to investigate the aforementioned SN properties $E_\mathrm{expl}$, $M_\mathrm{Ni}$ and $v_\mathrm{kick}$ separately for \casea, B and~C accretors, and single stars (Fig.~\ref{fig:Eexpl-MNi-vkick}). We find close relations between these properties and the central specific entropy $s_\mathrm{c}$ as indicated by Pearson's correlation coefficients $\rho$ in Fig.~\ref{fig:Eexpl-MNi-vkick}. The correlation of $s_\mathrm{c}$ and $M_\mathrm{Ni}$ is the tightest ($\rho>0.9$), followed by that of $s_\mathrm{c}$ with $E_\mathrm{expl}$ ($\rho\approx0.7\text{--}0.8$) and $v_\mathrm{kick}$ ($\rho\approx0.6\text{--}0.8$).

To help compare the SN outcomes of the different evolutionary histories of stars, \ie single stars and \casea, B and~C accretors, we fit quadratic functions to each of the entire data sets of $E_\mathrm{expl}$, $M_\mathrm{Ni}$ and $v_\mathrm{kick}$ and show these fits in Fig.~\ref{fig:Eexpl-MNi-vkick}. The root-mean-square deviation from the quadratic fits of $E_\mathrm{expl}$, $M_\mathrm{Ni}$ and $v_\mathrm{kick}$ are $0.4\times 10^{51}\,\mathrm{erg}$, $0.02\,\msun$ and $220\,\kms$, respectively. The found correlations are overall similar in the single stars and \casea, B and~C accretors, as indicated by the fit functions and the correlation coefficients $\rho$. Moreover, we also find no obvious differences between stellar models that evolve through a long-lived BSG phase \citepalias[\ie more than $0.2\,\mathrm{Myr}$ as a core-helium burning BSG as defined in][]{schneider2024a} and those that do not; similarly, the \caseah accretors with incomplete rejuvenation identified in \citetalias{schneider2024a} follow the same overall trend. This immediately implies that the past evolutionary histories of pre-SN stars do not matter much for the exact SN explosion outcome but rather only the pre-SN core structure as summarised here by the central specific entropy. This finding is in line with the relation between central entropy and gravitational NS mass mentioned above. Covering the same ranges of explosion properties does, of course, not mean that the population averages between these groups (single stars and the different accretor models) are also the same, and there can be systematic differences.

As argued in \citetalias{schneider2024a}, the central specific entropy $s_\mathrm{c}$, iron core mass $M_\mathrm{Fe}$ and compactness parameter $\xi_{2.5}$\footnote{The compactness $\xi_M$ measures the mass-to-radius ratio at a specific mass coordinate $M$ at the onset of core collapse, $\xi_M = (M/\msun)/(R(M)/1000\,\mathrm{km})$ \citep{oconnor2011a}. In this paper, we measure it at $M=2.5\,\msun$.} are all related to each other and are useful summary proxies of the core structures of our single stars and accretors for the explodability. Correlations of $E_\mathrm{expl}$, $M_\mathrm{Ni}$ and $v_\mathrm{kick}$ with the compactness parameter have also been noted by, \eg, \citet{nakamura2015a}, \citet{muller2016a}, \citet{schneider2021a} and \citet{burrows2024a}. The mass coordinate $M_4$ at a specific entropy of $s_\mathrm{c}/N_\mathrm{A} k_\mathrm{B}=4$ could also have been used instead of $s_\mathrm{c}$ and shows the same qualitative correlations. This value is also known to strongly correlate with the mass coordinate at which the SN shock is revived and hence the NS remnant mass as, \eg, explained in \citet{schneider2021a}. Overall, however, the correlations in this work are the strongest with central entropy $s_\mathrm{c}$. 

It is expected that $E_\mathrm{expl}$ and $v_\mathrm{kick}$ show a similar relation with $s_\mathrm{c}$, because $v_\mathrm{kick}$ is directly computed from $E_\mathrm{expl}$ in our SN model ($v_\mathrm{kick}\propto\sqrt{\Delta M\,E_\mathrm{expl}}/M_\mathrm{NS,grav}$, where $\Delta M$ is the neutrino heated mass). Moreover, The nickel mass $M_\mathrm{Ni}$ is also closely related to $E_\mathrm{expl}$ because $E_\mathrm{expl}$ sets the post-shock temperature in the SN ejecta which then determines which parts of the ejecta are hot enough to be explosively burnt into iron group elements, \ie nickel \citep[see Sect.~\ref{sec:nickel-proxy} below and, \eg,][]{lyman2016a, curtis2019a, suwa2019a, pejcha2020a, burrows2024a}. This very connection of explosion energy and nickel yield via the post-shock temperature is implemented in the \citet{muller2016a} SN code applied here and thus explains the found correlation. 

In conclusion, we find that the core properties of stars (\eg central entropy or compactness) mainly determine the explodability and the outcome of SN explosions in our models, and it is of lesser relevance which evolutionary pathway (\ie single star or \casea, B or~C accretor, and whether there was a long-lived BSG phase) led to a given core structure. A single parameter such as central entropy or compactness is insufficient to accurately predict the SN outcomes as is evident from the scatter in Fig.~\ref{fig:Eexpl-MNi-vkick} around the quadratic fits and even more well-known for the explodability of stars. It can nevertheless serve as a useful proxy.

\subsection{\label{sec:nickel-proxy}Nickel mass as a proxy for explosion energy and NS remnant mass}

\begin{figure*}
    \centering
    \includegraphics{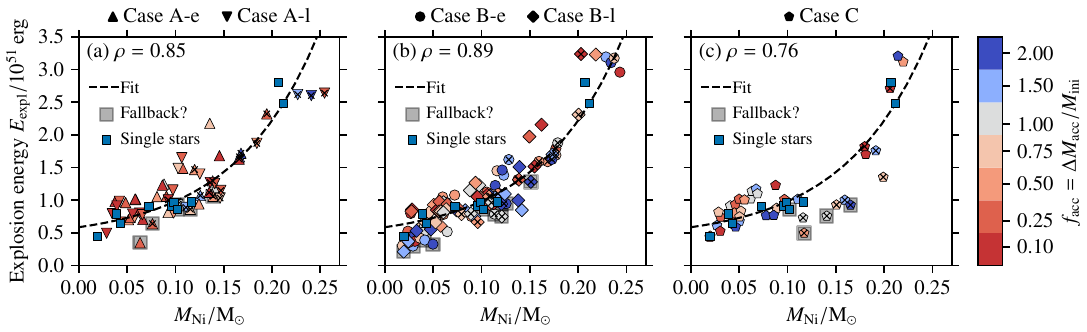}
    \caption{Nickel mass $M_\mathrm{Ni}$ as a function of explosion energy $E_\mathrm{expl}$. As in Fig.~\ref{fig:MNi_MNS}, Pearson's correlation coefficients $\rho$ are given as well as an exponential fit to the entire data set.}
    \label{fig:MNi-Eexpl}
\end{figure*}

\begin{figure*}
    \centering
    \includegraphics{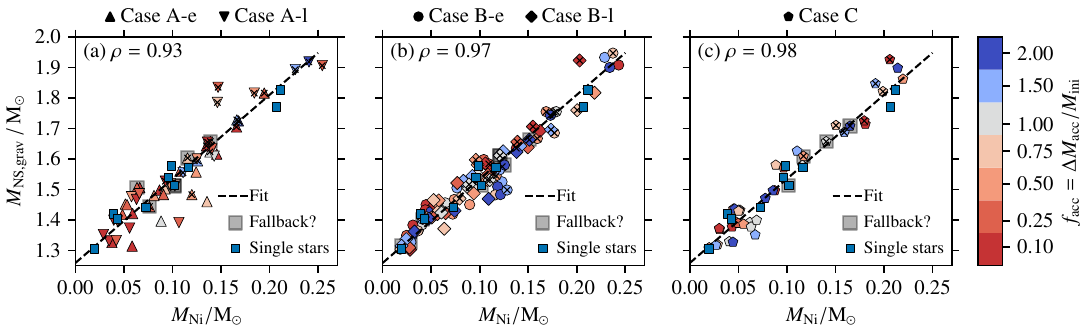}
    \caption{Nickel mass $M_\mathrm{Ni}$ as a function of NS mass $M_\mathrm{NS,grav}$. Pearson's correlation coefficients $\rho$ are provided, and a linear fit to the complete data set is shown for the \casea, B and~C accretors.}
    \label{fig:MNi_MNS}
\end{figure*}

Explosion energies and nickel yields are closely related, and this has been demonstrated by theoretical work as well as inferred from observations of SNe \citep[see description in Sect.~\ref{sec:explosion-properties} and, \eg, review by][]{pejcha2020a}. Indeed, we find a $M_\mathrm{Ni}$--$E_\mathrm{expl}$ relation that can be well fitted by an exponential function to the entire data set (Fig.~\ref{fig:MNi-Eexpl}),
\begin{align}
    \frac{E_\mathrm{expl}}{10^{51}\,\mathrm{erg}} & = (0.186\pm0.044) \exp\left[ (11.394\pm0.935) \frac{M_\mathrm{Ni}}{\msun} \right] \nonumber \\
    & \quad + (0.402\pm0.079),
    \label{eq:MNi-Eexpl-fit}
\end{align}
where the uncertainties are $1\sigma$ and purely statistical. The root-mean-square deviation of the entire model data from the fit function is $0.31\times 10^{51}\,\mathrm{erg}$. In Fig.~\ref{fig:MNi-Eexpl}, we again separate the sample into \casea, B and~C accretors, and show the single-star models in each panel to illustrate that the evolutionary history of stars does not seem to play a major role in setting this relation. Also, the models that experienced a long-lived BSG phase follow the same trend as do the \caseah systems with incomplete rejuvenation. In these relations, the SN fallback systems seem to deviate the strongest from the overall trend, and the scatter would be reduced if these systems were not taken into account.

Given the relations between $M_\mathrm{NS,grav}$ and $s_\mathrm{c}$ \citepalias[figure~13 in][]{schneider2024a}, and $M_\mathrm{Ni}$ and $s_\mathrm{c}$ (Fig.~\ref{fig:Eexpl-MNi-vkick}), it is evident that also $M_\mathrm{NS,grav}$ and $M_\mathrm{Ni}$ must be closely related. Indeed, we find a tight correlation between these quantities (Pearson's correlation coefficients of $\rho>0.93$; Fig.~\ref{fig:MNi_MNS}). From the least-squares fitting of the entire data set, we see that a linear relation,
\begin{align}
    \frac{M_\mathrm{NS,grav}}{\msun} = (2.757\pm0.045) \frac{M_\mathrm{Ni}}{\msun}  + (1.259\pm0.005),
    \label{eq:MNi-MNS-fit}
\end{align}
describes our models well (the purely statistical uncertainties are again $1\sigma$ and the root-mean-square deviation from the fit function is $0.045\,\msun$). As with all other relations found in this work, it is mostly independent of the evolutionary history of the stars. 

The fit suggests a minimum NS mass of $(1.259\pm0.005)\,\msun$ based on our stellar models and the applied SN explosions; this mass may still be larger than those of NSs from electron-capture SNe because such explosions are not considered here. It nevertheless seems to overlap with a possible ${\approx}\,1.25\,\msun$ NS component in the mass distribution of NSs suggested to be due to electron-capture SNe by \citet{schwab2010a}. There are also NSs with masses ${<}\,1.2\,\msun$ \citep{ozel2016a, tauris2017a, farrow2019a}. Our models do not make predictions on the maximum NS masses, but it is rather an assumption to help discriminate between NS and BH formation ($2\,\msun$ in this work, see Sect.~\ref{sec:methods}). Nevertheless, if it were possible to reliably calibrate Eq.~(\ref{eq:MNi-MNS-fit}), the maximum nickel mass observed in core-collapse SNe may put constraints on the maximum NS mass formed in such explosions. 

In terms of the physics of neutrino-driven explosions, one can understand this relation qualitatively as follows. Neutrino-driven explosions are a threshold process similar to line-driven stellar winds. Hence, they obey an explosion energy--binding energy relation \citep[see \eg][]{janka2017a, burrows2024a}. When models are more difficult to explode (\ie have high binding energy), the neutrino heating must be more intense until an explosion is triggered with a larger explosion energy. Models with high binding and explosion energies lead to larger nickel yields because of the $M_\mathrm{Ni}$--$E_\mathrm{expl}$ relation (Fig.~\ref{fig:MNi-Eexpl}). Moreover, the models with the largest binding energies also have the largest $M_\mathrm{Fe}$ and thus the largest $s_\mathrm{c}$ and $\xi_{2.5}$. $M_\mathrm{Fe}$ strongly correlates with $M_4$, the location where the SN explosion is revived, and hence the NS mass. Overall, it is therefore expected to find a $M_\mathrm{NS,grav}$--$M_\mathrm{Ni}$ correlation as shown in Fig.~\ref{fig:MNi_MNS} and Eq.~(\ref{eq:MNi-MNS-fit}).

In SN observations, the two relations in Eqs.~(\ref{eq:MNi-Eexpl-fit}) and~(\ref{eq:MNi-MNS-fit}) suggest that one can obtain the NS mass and the explosion energy by measuring the nickel mass. The latter may, \eg, be determined from the late-time, nickel-decay powered tails of SN light curves. In particular, the $M_\mathrm{Ni}$--$E_\mathrm{expl}$ relation may help break degeneracies in SN observations \citep[\cf][]{goldberg2019a}, but for any given $M_\mathrm{Ni}$ our models scatter by ${\approx}\,0.3\times10^{51}\,\mathrm{erg}$ around the best fit. A highly precise prediction of $E_\mathrm{expl}$ from a measured $M_\mathrm{Ni}$ thus seems challenging. For the NS masses, the scatter around the best fit is smaller, and a measured nickel mass may provide an NS mass within ${\approx}\,0.05\,\msun$. Moreover, the relation implies that there is a link between the minimum and maximum observed nickel masses and the minimum and maximum NS masses formed in neutrino-driven core-collapse SNe, respectively. However, at both ends of the nickel mass ranges, electron-capture SNe and explosions supplemented/powered by other SN mechanisms (\eg magnetar-driven SNe) will complicate the interpretation. While our relations may give a qualitative picture and clearly show that there are $M_\mathrm{Ni}$--$M_\mathrm{NS,grav}$ and $M_\mathrm{Ni}$--$E_\mathrm{expl}$ relations, Eqs.~(\ref{eq:MNi-Eexpl-fit}) and~(\ref{eq:MNi-MNS-fit}) depend quantitatively on the calibrations of our SN engine and the stellar models.

\subsection{\label{sec:hrd}Hertzsprung--Russell diagram and SN classification}

\begin{figure*}
    \centering
    \includegraphics{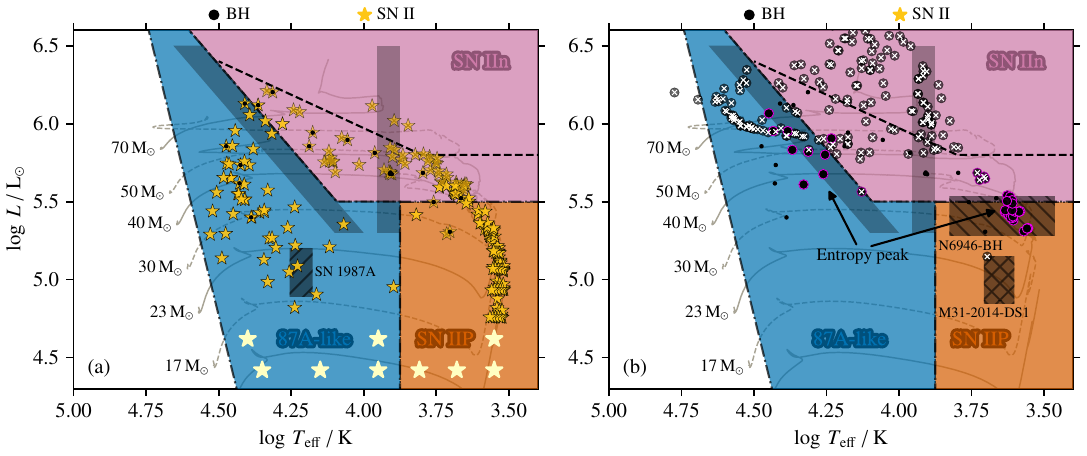}
    \caption{Hertzsprung--Russell diagram of the pre-SN locations of the single star and accretor models. In the left panel (a), we show models successfully exploding in SNe, while we show BH formation in the right panel (b). Small black circles indicate BH formation by SN fallback and the magenta contours highlight cases of BH formation from stars in the entropy peak at $M_\mathrm{CO}=6\text{--}8\,\msun$ (\cf Fig.~\ref{fig:MCO-entropy}). The dark grey, hatched areas show the locations of the BSG progenitor of \snesa (left panel only) and the CSGs N6946-BH and M31-2014-DS1 that have been suggested to have collapsed into a BH (right panel only). The hot and cool regimes of the S~Doradus instability strip are indicated by the light grey bands \citep{smith2004a} and the Humphreys-Davidson limit by the black dashed line. Models slightly faded out and additionally marked by white and golden crosses spend more than $10^4\,\mathrm{yr}$ in the S~Doradus region and thus likely experience significant mass loss (\ie they will most likely explode elsewhere in the HR diagram). The reddish, blue and purple shaded areas show our ad-hoc SN classification into \sniip, \snesalike and interacting \sniin, respectively (see main text for details). An ordering of the models into Case~A, B and C accretors and the individual accretion fractions can be found in the HR diagrams in Figure~9 of \citetalias{schneider2024a}.}
    \label{fig:hrd-all}
\end{figure*}

In \citetalias{schneider2024a}, we have discussed the pre-SN positions of our single star and accretor models in the HR diagram. Using these positions, we group the models into three broad SN classes (Fig.~\ref{fig:hrd-all}).

\paragraph*{\sniin:} 
This group shall represent the broad class of interacting SNe (including interacting superluminous SNe, SLSNe). To define potential interacting SNe, we consider SN progenitors that explode as LBVs, \ie as stars inside the S~Doradus instability strip of the HR diagram \citep{smith2004a} or as cooler super/hypergiants ($T_\mathrm{eff}<7500\,\mathrm{K}$) with luminosities of $\log L/\lsun>5.5$ (\cf pink coloured region in Fig.~\ref{fig:hrd-all}). The latter is the empirically found luminosity limit of RSGs in M31, the Milky Way and the Large and Small Magellanic Clouds \citep{davies2018b, mcdonald2022a}. Both criteria reflect the idea that such stars may have lost significant amounts of mass just before the SN such that the SN ejecta can interact with the lost mass in a way that affects the SN light curve. Alternatively, some stars might lose a significant fraction of their envelope mass such that they could instead explode as \sneibc. When stars spend more than $10^4\,\mathrm{yr}$ in this region of the HR diagram, we flag them up throughout this paper as stars with likely significant/enhanced LBV-like mass loss (often indicated by cross symbols in figures). This is to indicate that the properties of these stars might be quite different from what is shown here because our models do not apply LBV-like mass loss\footnote{The choice of the duration of $10^4\,\mathrm{yr}$ in the LBV region for flagging up models is ad-hoc and guided by the average mass loss rates of LBVs, which is of the order of $10^{4}\,\msun\,\mathrm{yr}^{-1}$ \citep[see, \eg,][]{smith2014a}. In such cases, about $\mathcal{O}(1\,\msun)$ is lost and could thus significantly affect the appearance of the stars. In the most extreme cases, the entire hydrogen-rich envelope could be lost. In our models, stars generally do not evolve out of the LBV regime once they enter it. Hence, the time spent as an LBV in almost all of our models implies they spent this time in the LBV regime before core-collapse.}.

\paragraph*{\sniip:} 
We denote SNe from hydrogen-rich progenitor stars with luminosities of $\log L/\lsun\leq5.5$ and $T_\mathrm{eff}\leq7500\,\mathrm{K}$ as \sneiip. The effective temperature limit is the approximate cool side of the S~Doradus instability strip. Not all hydrogen-rich SN progenitors that explode in this part of the HR diagram will give rise to a \sniip, but some may rather be classified as, \eg, \sneiil or \sneiib \citep[SN explosions in different phases of RSG pulsations can also affect the light-curve classification as \sniip and \sniil, see \eg][]{bronner2025a}. Type~IIb SNe have hydrogen-rich envelope masses of ${\lesssim}1.0\,\msun$ \citep[][]{yoon2017b, sravan2019a} and are not found in our accretor models, which, by construction, have relatively massive hydrogen-rich envelopes.

\paragraph*{\snesalike:}
Hydrogen-rich SN progenitors that are on the hot side of the S~Doradus instability strip or that are hotter than $7500\,\mathrm{K}$ for $\log L/\lsun\leq5.5$ may produce light curves similar to that of \snesa (foremost, they may show long-rising, peculiar SN~II light curves; \cf Fig.~\ref{fig:hrd-all}).

\vspace*{0.5cm}
While these classifications cannot be accurate and cannot reflect the large diversity of SN~II, they serve as a first way to connect our pre-SN models to their likely appearance as SNe and thus offer a more direct way to compare to observations of SNe.

In the HR diagram, we see that our models cover all three SN groups (Fig.~\ref{fig:hrd-all}). The lowest initial mass in our grid is $11\,\msun$ and we thus lack pre-SN models at $\log L/\lsun \lesssim 4.7$. To indicate that there should be many more SN progenitors at these luminosities, we add star symbols to the bottom of the HR diagram. Moreover, we can also see that most models in the \sniin group spend more than $10^4\,\mathrm{yr}$ in this part of the HR diagram and only a few evolve into it less than $10^4\,\mathrm{yr}$ before core collapse.

\begin{figure}
    \centering
    \includegraphics{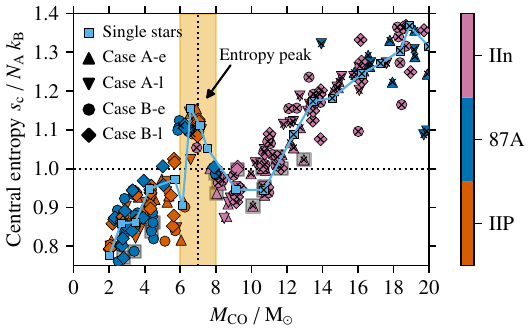}
    \caption{Central specific entropy $s_\mathrm{c}$ as a function of CO core mass $M_\mathrm{CO}$ at core collapse for the single-star and \casea and~B accretor models. The \casec accretors are not shown to enhance the clarity of the diagram. The entropy peak at $M_\mathrm{CO}\approx 6\text{--}8\,\msun$ associated with BH formation is indicated by the yellow-ish box. Colours relate the models to their pre-SN location in the HR diagram and hence their likely appearance as \sniin, \sniip and \snesalike. The horizontal and vertical black dotted lines help to visually distinguish between the different SN groups and the typical central entropies that models in these groups have at core collapse. The black crosses indicate models that likely experience significant LBV-like mass loss, and grey boxes show cases of likely SN fallback.}
    \label{fig:MCO-entropy}
\end{figure}

As shown in \citetalias{schneider2024a}, the compactness/entropy as a function of $M_\mathrm{CO}$ has a prominent peak at $M_\mathrm{CO}\approx7\,\msun$ in our models (Fig.~\ref{fig:MCO-entropy}). This peak is characteristic of all of our stellar models and regardless of their evolutionary histories. It is found in single stars of varying convective core boundary mixing parameters \citep{temaj2024a}, binary-stripped stars at two different metallicities \citep{schneider2021a, schneider2023a}, and also accretors of binary mass transfer and stellar mergers \citepalias{schneider2024a}. The peak is a consequence of core carbon burning becoming neutrino-dominated as shown and explained by \citet{laplace2025a}. It is also seen in stellar models computed with other codes and employing different physics assumptions \citep[\eg][]{oconnor2011a, sukhbold2014a, sukhbold2018a, chieffi2020a, takahashi2023a}. 

The models in the compactness/entropy peak form BHs in our models \citepalias{schneider2024a}. These BH-forming models roughly follow the hot side of the S~Doradus instability strip in the HR diagram for stars that reach core collapse as a BSG, and at a luminosity of $\log\,L/\lsun \approx 5.4$ for stars collapsing as RSGs. This implies that our groups of \snesa and \sniip are almost exclusively from stars with $M_\mathrm{CO}\lesssim7\,\msun$\footnote{Some very massive stars with $M_\mathrm{CO}\gtrsim14\,\msun$ lost so much mass that they are BSGs at core collapse and thereby enter the \snesalike group. However, these stars then have very low hydrogen-rich envelope masses.} while interacting SNe (\sneiin) have $M_\mathrm{CO}\gtrsim7\,\msun$ (Fig.~\ref{fig:MCO-entropy}). As shown in Sect.~\ref{sec:explosion-properties}, the central entropy can be regarded as a proxy for the SN outcome in terms of, \eg, explosion energy, nickel yield and SN kick velocity. Because the groups of \snesalike and \sniip models cover a similar range of central entropies (Fig.~\ref{fig:MCO-entropy}), they are expected to have similar explosion properties but possibly very different ejecta masses and radii. 

As explained in \citetalias{schneider2024a}, both the single stars and binary accretors that do not become long-lived BSGs explode along a similar band in the HR diagram (Fig.~\ref{fig:hrd-all}). Moreover, these models all follow the same CO core mass--luminosity relation such that the compactness/entropy peak of pre-core-collapse CSGs/RSGs is at a similar luminosity of $\log\,L/\lsun\,{\approx}\,5.4$ in the HR diagram. This model prediction coincides nicely with the inferred luminosity of the RSG N6946-BH that has been suggested to have collapsed into a BH \citep[Fig.~\ref{fig:hrd-all}; see \eg][but also JWST observations of \citealt{beasor2024a, kochanek2024a} that may challenge this interpretation]{gerke2015a, adams2017a, adams2017b, basinger2021a}. M31-2014-DS1 is another CSG that suddenly disappeared and that may have collapsed into a BH \citep{de2024a}. Its inferred luminosity of $\log L/\lsun \approx 5.0$\footnote{\citet{de2024a} do not provide an uncertainty to their inferred luminosity and we have assumed $\pm0.15\,\mathrm{dex}$ to indicate the position of this CSG in the HR diagram in Fig.~\ref{fig:hrd-all}.} does not match our prediction of BH-forming CSGs in the compactness/entropy peak. However, \citet{de2024a} suggest that the progenitor was a star that lost most of its hydrogen-rich envelope. The fast loss of parts of an envelope can make stars drop in surface luminosity, and there is a model in our grid that should have undergone additional mass loss from LBV-like winds and has dropped in luminosity before core collapse (Fig.~\ref{fig:hrd-all}). Luminosities of cool and red supergiants are notoriously uncertain, in particular if only limited information is available, as is the case for many SN progenitors \citep[see \eg][]{beasor2024b}. This is exemplified by the progenitor of SN~2023ixf that has decent observational coverage in multiple bands and still the bolometric luminosities inferred for the progenitor by different groups scatter easily over a range of ${\approx}\,0.6\,\mathrm{dex}$ \citep[\eg][]{ransome2024a}.

On average, the models classified as \sniin have larger central entropies than our \snesa and \sniip models (Fig.~\ref{fig:MCO-entropy}). Moreover, they also originate on average from larger initial masses as they have larger $M_\mathrm{CO}$. While there is no universal relation between the initial masses and $M_\mathrm{CO}$ in our accretor and single-star models (in particular for \casea accretors), this relation is quite tight in \caseb and~C accretors and similar to that of single stars (deviations in $M_\mathrm{CO}$ of accretor and single-star models on the order of 10\%, \cf figure~5 in \citetalias{schneider2024a}). The latter implies that the SN rate of our \sniin models is predicted to be lower compared to those of \snesalike and \sneiip.

\subsection{\label{sec:sn-diversity}Supernova diversity}

For easier discussion of the diversity of SNe and also as an overview of all our models, we summarise the ranges of various properties of the SNe and their progenitor stars in Table~\ref{tab:sn-properties} for our single-star models and the different SN groups defined in Sect.~\ref{sec:hrd}. We will refer back to this table throughout the rest of this paper. 

\begin{table}
    \caption{\label{tab:sn-properties}Ranges of selected properties of SNe and their progenitor stars.}
    \centering
    \addtolength{\tabcolsep}{-5.0pt}
    \begin{tabular}{lcccc}
        \toprule
         & Single stars & \sniip & \snesalikeshort & \sniin \\
        \midrule
        $s_\mathrm{c}/N_\mathrm{A} k_\mathrm{B}$ & $0.78\text{--}1.09$ & $0.78\text{--}1.10$ & $0.76\text{--}1.10$ & $0.87\text{--}1.17$ \\
        & ($0.78\text{--}1.05$) & ($0.78\text{--}1.10$) & ($0.76\text{--}1.10$) & ($0.87\text{--}1.03$) \\
        $\xi_{2.5}$ & $0.01\text{--}0.44$ & $0.01\text{--}0.51$ & $0.02\text{--}0.52$ & $0.17\text{--}0.52$ \\
        & ($0.01\text{--}0.40$) & ($0.01\text{--}0.51$) & ($0.02\text{--}0.52$) & ($0.17\text{--}0.36$) \\
        $M_\mathrm{CO}/\msun$ & $2.0\text{--}12.4$ & $2.0\text{--}12.1$ & $2.0\text{--}8.0$ & $6.7\text{--}15.1$ \\
        & ($2.0\text{--}9.1$) & ($2.0\text{--}7.0$) & ($2.0\text{--}8.0$) & ($7.2\text{--}9.2$) \\
        $M_\mathrm{cc}/\msun$ & $10.0\text{--}16.0$ & $11.0\text{--}57.1$ & $18.3\text{--}85.5$ & $13.0\text{--}99.5$ \\
        & ($10.0\text{--}13.9$) & ($11.0\text{--}33.6$) & ($18.3\text{--}85.5$) & ($13.0\text{--}85.9$) \\
        $M_\mathrm{ej}/\msun$ & $8.6\text{--}13.8$ & $9.6\text{--}55.1$ & $16.7\text{--}83.7$ & $11.3\text{--}97.6$ \\
        & ($8.6\text{--}12.1$) & ($9.6\text{--}32.0$) & ($16.7\text{--}83.7$) & ($11.3\text{--}84.1$) \\
        $M_\mathrm{H,env}/\msun$ & $1.2\text{--}7.2$ & $4.9\text{--}44.9$ & $14.6\text{--}75.1$ & $1.2\text{--}85.8$ \\
        & ($2.6\text{--}7.2$) & ($4.9\text{--}26.2$) & ($14.6\text{--}75.1$) & ($2.0\text{--}74.1$) \\
        $R_\mathrm{cc}/\rsun$ & $150\text{--}1050$ & $400\text{--}1250$ & $50\text{--}200$ & $100\text{--}1100$ \\
        & ($750\text{--}1050$) & ($400\text{--}1250$) & ($50\text{--}200$) & ($100\text{--}1050$) \\
        $T_\mathrm{eff,cc}\,/\,\mathrm{kK}$ & $3.3\text{--}13.7$ & $3.3\text{--}6.6$ & $7.9\text{--}33.8$ & $4.3\text{--}23.1$ \\
        & ($3.3\text{--}5.2$) & ($3.3\text{--}6.6$) & ($7.9\text{--}33.8$) & ($4.3\text{--}23.1$) \\
        $\log\,L_\mathrm{cc}\,/\,\lsun$ & $4.8\text{--}5.8$ & $4.8\text{--}5.5$ & $4.8\text{--}6.1$ & $5.5\text{--}6.2$ \\
        & ($4.8\text{--}5.6$) & ($4.8\text{--}5.4$) & ($4.8\text{--}6.1$) & ($5.5\text{--}6.1$) \\
        $E_\mathrm{expl}/10^{51}\,\mathrm{erg}$ & $0.4\text{--}2.8$ & $0.4\text{--}3.2$ & $0.2\text{--}3.2$ & $0.4\text{--}3.2$ \\
        & ($0.4\text{--}2.8$) & ($0.4\text{--}3.2$) & ($0.2\text{--}3.2$) & ($0.4\text{--}1.7$) \\
        $M_\mathrm{Ni}/\msun$ & $0.02\text{--}0.21$ & $0.02\text{--}0.24$ & $0.02\text{--}0.23$ & $0.06\text{--}0.25$ \\
        & ($0.02\text{--}0.21$) & ($0.02\text{--}0.24$) & ($0.02\text{--}0.23$) & ($0.06\text{--}0.18$) \\
        $v_\mathrm{kick}/\kms$ & $250\text{--}1450$ & $250\text{--}1650$ & $150\text{--}1600$ & $250\text{--}1900$ \\
        & ($250\text{--}1450$) & ($250\text{--}1650$) & ($150\text{--}1600$) & ($250\text{--}1200$) \\
        $\log E_\mathrm{expl}/M_\mathrm{ej}$ & $16.4\text{--}17.1$ & $16.0\text{--}17.1$ & $15.5\text{--}16.5$ & $15.8\text{--}17.0$ \\
        & ($16.4\text{--}17.1$) & ($16.0\text{--}17.1$) & ($15.5\text{--}16.5$) & ($15.8\text{--}16.8$) \\
        $v_\mathrm{sn}/\kms$ & $2300\text{--}5100$ & $1400\text{--}4900$ & $800\text{--}2600$ & $1100\text{--}4600$ \\
        & ($2300\text{--}5100$) & ($1500\text{--}4900$) & ($800\text{--}2600$) & ($1100\text{--}3700$) \\
        \bottomrule
    \end{tabular}
    \addtolength{\tabcolsep}{+1.0pt}
    \tablefoot{The \sniip, \snesalike and \sniin categories only include our accretor models, and the reported ranges only differ in a few quantities when neglecting models that likely experience enhanced LBV-like mass loss (modified ranges shown in parentheses). Provided are the central specific entropy $s_\mathrm{c}$ of stars at core collapse, the compactness $\xi_{2.5}$, the CO core masses $M_\mathrm{CO}$, the final stellar masses $M_\mathrm{cc}$, the SN ejecta masses $M_\mathrm{ej}$, the mass of the hydrogen-rich envelope $M_\mathrm{H,env}$ of the SN progenitors, the radii $R_\mathrm{cc}$, the effective temperature $T_\mathrm{eff,cc}$ and luminosity $\log\,L_\mathrm{cc}/\lsun$ of the pre-SN models, the explosion energy $E_\mathrm{expl}$, nickel mass $M_\mathrm{Ni}$, NS kick velocity $v_\mathrm{kick}$, logarithmic $E_\mathrm{expl}$ to $M_\mathrm{ej}$ ratio (in units of $\mathrm{erg}\,\mathrm{g}^{-1}$) and the average ejecta velocity $v_\mathrm{sn}=\sqrt{2E_\mathrm{expl}/M_\mathrm{ej}}$. All numbers are rounded to the last shown digit except for $R_\mathrm{cc}$, $v_\mathrm{kick}$ and $v_\mathrm{sn}$, which are rounded to the nearest $50\,\rsun$, $50\,\kms$ and $100\,\kms$, respectively. The value ranges in parentheses exclude stars that likely experience significant LBV-like mass loss, \ie spend more than $10^4\,\mathrm{yr}$ as LBVs. For CO core masses around $\approx7\,\msun$, there are no SN explosions because these models are predicted to directly collapse into BHs \citepalias[\cf][]{schneider2024a}. For individual numbers and further quantities, see Table~\ref{tab:models}.}
\end{table}

\begin{figure*}
    \centering
    \includegraphics{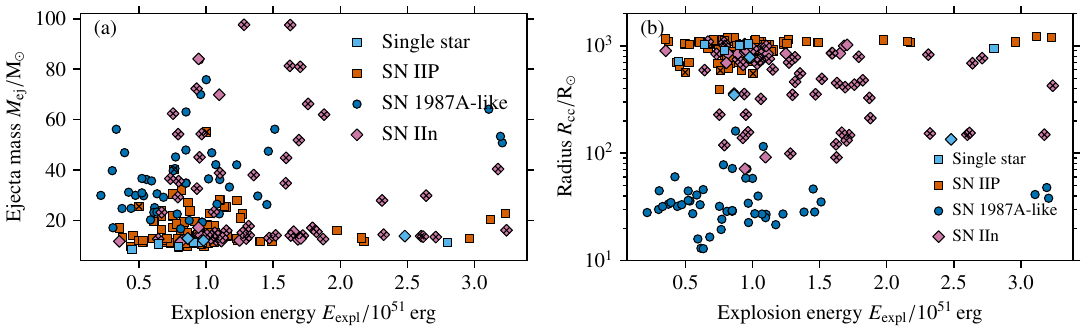}
    \caption{Ejecta mass $M_\mathrm{ej}$ (a) and SN progenitor radius $R_\mathrm{cc}$ (b) as a function of the SN explosion energy $E_\mathrm{expl}$. The symbols indicate our rough SN classification based on the position of the SN progenitors in the HR diagram. Single-star models are shown in light blue to distinguish them from the accretor models. As before, black crosses indicate models likely experiencing large LBV-like mass loss.}
    \label{fig:Eexpl-Mej-R}
\end{figure*}

As shown in Sect.~\ref{sec:explosion-properties}, the different evolutionary histories studied in this work (single star evolution vs.\ \casea, B and~C accretors) can lead to similar pre-SN core structures summarised, \eg, by the central entropy or compactness that then results in similar SN engines (\ie explosion energies, produced nickel mass and NS kick velocity). While these explosion properties may be similar, the envelope structures of the SN progenitors can be very different. In Fig.~\ref{fig:Eexpl-Mej-R} (see also Table~\ref{tab:sn-properties}), we show the SN ejecta mass $M_\mathrm{ej}$ and SN progenitor radius $R_\mathrm{cc}$ as a function of the explosion energy $E_\mathrm{expl}$\footnote{One could have also replaced the explosion energy by the ejected nickel mass or NS kick velocity and would still get the qualitatively same picture.}. We can see an enhanced scatter and entirely different branches due to the accretor models that contribute to the observed SN diversity.

The applied stellar wind mass loss restricts the final masses of our single-star, solar-like metallicity models to values of ${<}\,16\,\msun$ and thus limits their ejecta masses to ${<}\,14\,\msun$ (Fig.~\ref{fig:Eexpl-Mej-R}a and Table~\ref{tab:sn-properties}). In contrast, the accretor models can greatly exceed this if they spend considerable time as a BSG where stellar winds in our models are the weakest \citepalias[see also][]{schneider2024a}. While some of the \snesalike and \sniin accretor models spend all their post-accretion evolution as BSGs, the \sniip progenitors can -- by definition -- not do that. Hence, they must have systematically smaller maximum ejecta masses as is evident from Fig.~\ref{fig:Eexpl-Mej-R}a and Table~\ref{tab:sn-properties}. To form a long-lived BSG progenitor for \snesalikeshort SNe, the accretor must accrete beyond a certain mass threshold \citepalias[see][and references therein]{schneider2024a}. Hence, such BSG SN progenitors must have a systematically larger minimum ejecta mass compared to both \sneiip and \sneiin. Lastly, \sneiin are generally from stars with larger masses (\eg as indicated by their larger CO core masses, Table~\ref{tab:sn-properties}) such that they also have the highest ejecta masses in our models (Fig.~\ref{fig:Eexpl-Mej-R}a). Note, however, that our models do not account for the likely encountered LBV-like winds and that our final stellar and thus ejecta masses are upper limits. Our models are for solar-like metallicity, and the final and hence ejecta masses are expected to be larger (smaller) at a lower (higher) metallicity. The systematic differences between the SN classes are thought to persist also at other metallicities. 

The radii of our BSG \snesalikeshort SN progenitors are the smallest in our models ($R_\mathrm{cc} \approx 50\text{--}200\,\rsun$; Fig.~\ref{fig:Eexpl-Mej-R}b and Table~\ref{tab:sn-properties}). The largest radii are found in our single stars and \sniip progenitors (a few $100$ to $1000\,\rsun$) while those of \sneiin populate the radii in between and overlap with those of the single stars and \sniip progenitors (Fig.~\ref{fig:Eexpl-Mej-R}b). The radii of SN progenitors directly influence the appearance of SNe. For example, the luminosities of SN shock breakout and the general SN light curve are (partly) related to the progenitors' radii, with smaller radii leading to lower luminosities and vice versa (\cf also the plateau luminosity of \sneiip). Moreover, the radii already indicate another important difference in the envelope structures of stars: more compact stars have radiative envelopes while those of larger stars are convective. Radiative and convective envelopes have quite different density profiles that are expected to directly influence the shapes of SN light curves. 

Lastly, the ejecta masses of the SN progenitors of our accretor models are larger than in the single-star models and, in particular, those of \snesalike progenitors have the systematically highest minimum ejecta masses as explained above. A large ejecta mass immediately implies a long photon diffusion timescale and hence a slowly evolving SN light curve. We will discuss this more quantitatively for \sneiip in Sect.~\ref{sec:light curves}. For our \snesalike progenitors, we predict the slowest evolution to maximum light given the likely largest ejecta masses (especially the largest minimum ejecta masses). Our \sneiin may also evolve similarly slowly in certain cases and still will depend strongly on the uncertain final stellar mass. Hence, these structural differences in SN progenitors can help understand the observed large diversity of hydrogen-rich SNe.

\begin{figure}
    \centering
    \includegraphics{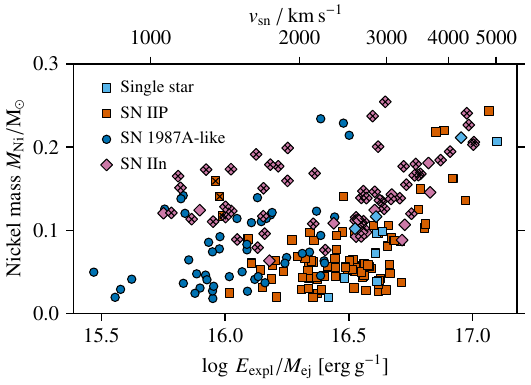}
    \caption{Similar to Fig.~\ref{fig:Eexpl-Mej-R} but showing the ejected nickel mass $M_\mathrm{Ni}$ as a function of logarithmic explosion-energy to ejecta-mass ratio $E_\mathrm{expl}/M_\mathrm{ej}$. The $E_\mathrm{expl}/M_\mathrm{ej}$ ratio is a measure of the average SN ejecta velocity, $v_\mathrm{sn} = \sqrt{2E_\mathrm{expl}/M_\mathrm{ej}}$, shown on the top $x$-axis. Black crosses are for models likely experiencing significant LBV-like mass loss.}
    \label{fig:EM-MNi}
\end{figure}

From the spectroscopy of SNe, it is possible to measure the expansion speed at different epochs. We cannot predict a time evolution of the SN explosions from our models but we can compute the explosion energy--ejecta mass ratio $E_\mathrm{expl}/M_\mathrm{ej}$ which is related to the average ejecta velocity $v_\mathrm{sn} = \sqrt{2 E_\mathrm{expl}/M_\mathrm{ej}}$. While the range of explosion energies in all our models (single stars as well as all accretors) is similar, the range of ejecta masses is very different (Table~\ref{tab:sn-properties}). Hence, also the ranges of $E_\mathrm{expl}/M_\mathrm{ej}$ and $v_\mathrm{sn}$ are different as illustrated in Fig.~\ref{fig:EM-MNi}. The \sniip and \sniin accretors overlap with the single stars in terms of $E_\mathrm{expl}/M_\mathrm{ej}$ in Fig.~\ref{fig:EM-MNi} but are more broadly scattered and extend to lower values because of the larger range in ejecta masses. Moreover, our \sneiin have the highest minimum nickel mass and are thus offset with respect to all other models. There is hardly any overlap of the \snesalike accretor models and single-star progenitors in the $E_\mathrm{expl}/M_\mathrm{ej}$--$M_\mathrm{Ni}$ plane because of distinctly different ejecta masses (Fig.~\ref{fig:EM-MNi}).

\subsection{\label{sec:light curves}\sniip light curves}

The light curves of \sneiip are characterised by an almost constant luminosity, called the plateau luminosity $L_\mathrm{p}$, that is maintained for a relatively long time $t_\mathrm{p}$ of order $100\,\mathrm{d}$. Several analytic scaling relations for $L_\mathrm{p}$ and $t_\mathrm{p}$ have been derived \citep[\eg][]{arnett1980a, chugai1991a, popov1993a, kasen2009a}, and such analytic and more general functions were fitted to computed light curves of \sneiip \citep[\eg][]{kasen2009a, sukhbold2016a, goldberg2019a, fang2025a}. Here, we use the fit formulae obtained by \citet{goldberg2019a} to estimate the plateau luminosity $L_\mathrm{p,50}$, the duration of the plateau $t_\mathrm{p}$ and the velocity of the Fe~II $\lambda5169$ line  $v_\mathrm{p,50}$, which traces the SN photosphere\footnote{While the Fe~II $\lambda5169$ line traces the photosphere, its velocity is systematically higher than that of the photosphere. We will come back to this difference later in this section.} and is readily observable, for \sniip-like explosions from our model progenitors,
\begin{align}
    &\log \left(\frac{L_\mathrm{p,50}}{10^{42}\,\mathrm{erg}\,\mathrm{s^{-1}}}\right) = 42.16 - 0.40 \log \left(\frac{M_\mathrm{ej}}{10\,\msun}\right) \nonumber\\ 
    &\quad\quad\quad + 0.74 \log \left(\frac{E_\mathrm{expl}}{10^{51}\,\mathrm{erg}}\right) + 0.76 \log \left(\frac{R_\mathrm{cc}}{500\,\rsun}\right),\label{eq:Lp}\\
    &\log \left(\frac{t_\mathrm{p}}{\mathrm{d}}\right) = 2.184 + 0.134 \log \left(\frac{M_\mathrm{Ni}}{\msun}\right) \nonumber\\
    &\quad\quad\quad + 0.411 \log \left(\frac{M_\mathrm{ej}}{10\,\msun}\right) - 0.282 \log \left(\frac{E_\mathrm{expl}}{10^{51}\,\mathrm{erg}}\right),\label{eq:tp}\\
    &\log \left(\frac{v_\mathrm{Fe,50}}{\kms}\right) = 3.61 - 0.12 \log \left(\frac{M_\mathrm{ej}}{10\,\msun}\right) \nonumber\\
    &\quad\quad\quad + 0.30 \log \left(\frac{E_\mathrm{expl}}{10^{51}\,\mathrm{erg}}\right) + 0.25 \log \left(\frac{R_\mathrm{cc}}{500\,\rsun}\right) \label{eq:vFe}.
\end{align}
The plateau luminosity and Fe-line velocity are evaluated at $50\,\mathrm{d}$ after shock-breakout. These fit functions are in good agreement with the expected analytic scaling relations of \citet{popov1993a} and the extensions thereof by \citet{kasen2009a}. \sneiip are associated with RSG progenitors \citep[see \eg][]{smartt2009a, smartt2015a}, and we thus compute $L_\mathrm{p,50}$, $t_\mathrm{p}$ and $v_\mathrm{Fe,50}$ only for our group of \sniip models. The relations are scaled to a typical RSG SN progenitor with $R_\mathrm{cc}\,{=}\,500\,\rsun$, $E_\mathrm{expl}\,{=}\,10^{51}\,\mathrm{erg}$ and $M_\mathrm{ej}\,{=}\,10\,\msun$.

\begin{figure}
    \centering
    \includegraphics{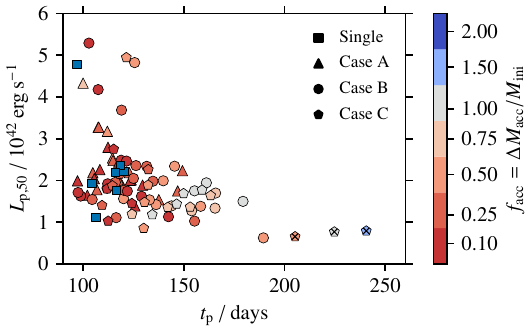}
    \caption{Plateau luminosity $L_\mathrm{p,50}$ at 50\,d after shock breakout and duration $t_\mathrm{p}$ of \sniip light curves for our single-star and accretor models following the scaling relations of \citet{goldberg2019a}. Black crosses show models with likely large LBV-like mass loss.}
    \label{fig:sn-light curves}
\end{figure}

We show $L_\mathrm{p,50}$ as a function of $t_\mathrm{p}$ in Fig.~\ref{fig:sn-light curves}. While there is little variation in the plateau duration $t_\mathrm{p}$ of our single star models (${\approx}\,100\text{--}125\,\mathrm{d}$), the \sniip light curves from our exploding RSG progenitors with a mass accretion history can be longer by a factor of 2 (we find values of up to ${\approx}\,250\,\mathrm{d}$). The longer plateau duration is caused by the higher envelope and thus ejecta masses of these stars \citepalias[see Sect.~\ref{sec:sn-diversity} and Table~\ref{tab:sn-properties}, and also][]{schneider2024a}. The higher ejecta masses also reduce the plateau luminosity by a similar factor (\cf Eqs.~\ref{eq:Lp} and~\ref{eq:tp}), and the general trend of longer plateau durations and lower plateau luminosities with higher accretion fractions is evident in Fig.~\ref{fig:sn-light curves}. Both our single-star and accretor models cover a similar range of plateau luminosities, and most models cluster at $L_\mathrm{p,50}\approx 1\text{--}3\times 10^{42}\,\mathrm{erg\,s^{-1}}$. Similarly, most plateau durations are found at $t_\mathrm{p} \approx 100\text{--}175\,\mathrm{d}$.

In principle, accretors and stellar mergers can have helium-enriched envelopes compared to single stars, which we do not model here. The different helium content modifies the opacity of SN ejecta and thereby affects $L_\mathrm{p}$ and $t_\mathrm{p}$. A higher helium content reduces the electron scattering opacity and thus increases $L_\mathrm{p}$ and decreases $t_\mathrm{p}$ \citep[\cf][]{popov1993a, kasen2009a, goldberg2019a}. While this counteracts the longer plateau durations and lower luminosities from the higher ejecta masses in helium-enriched accretor models, the magnitude of this effect is less compared to that due to varying ejecta masses: For electron scattering, the opacity can change at most by a factor of 2 between pure hydrogen and pure helium plasmas. The opacity enters $L_\mathrm{p}$ and $t_\mathrm{p}$ with power-law exponents of $-1/3$ and $1/6$ \citep{popov1993a}, respectively, such that $L_\mathrm{p}$ at most reduces by 20\% and $t_\mathrm{p}$ increases at most by 10\%.

\sneiip are used for cosmological distance determinations, similarly to Type Ia SNe. There are two main techniques, the expanding photosphere method \citep{kirshner1974a} and the standardised candle method \citep[SCM;][]{hamuy2002a}. While the former is independent of calibrations and thus provides a direct method for distance determinations, the latter technique is based on an empirically calibrated relation between the expansion velocity and the plateau luminosity of Type~IIP SNe. Recent applications of this and related methods \citep[\eg][]{maguire2010a, rodriguez2014a, gall2018a, rodriguez2019a, dejaeger2020a, dejaeger2022a} have been successfully applied to Type~IIP SNe and, \eg, are used in investigating the tension in the Hubble constant inferred from local and distant measurements \citep[\eg][]{dejaeger2023a}.

The plateau luminosity from the applied fitted relations of \citet[][Eq.~\ref{eq:Lp}]{goldberg2019a} has the theoretically-expected scaling of $L_\mathrm{p,50}\propto (v_\mathrm{ph,50})^2$ \citep[see \eg][]{kasen2009a} with the photospheric velocity $50\,\mathrm{d}$ after shock-breakout,
\begin{align}
    &\log \left(\frac{v_\mathrm{ph,50}}{\kms}\right) = 3.54 - 0.19 \log \left(\frac{M_\mathrm{ej}}{10\,\msun}\right) \nonumber\\
    &\quad\quad\quad + 0.36 \log \left(\frac{E_\mathrm{expl}}{10^{51}\,\mathrm{erg}}\right) + 0.32 \log \left(\frac{R_\mathrm{cc}}{500\,\rsun}\right). \label{eq:vph}
\end{align}
Observationally, one rather measures the systematically higher velocity $v_\mathrm{Fe,50}$ that has very similar scalings with $M_\mathrm{ej}$, $E_\mathrm{expl}$ and $R_\mathrm{cc}$ as the photospheric velocity (\cf Eqs.~\ref{eq:vFe} and~\ref{eq:vph}). Hence, $L_\mathrm{p,50}$ and $v_\mathrm{Fe,50}$ follow an almost one-to-one relation in our models. Moreover, \citet{goldberg2019a} argue that this is the fundamental reason why it is impossible to break the degeneracy between $M_\mathrm{ej}$, $E_\mathrm{expl}$ and $R_\mathrm{cc}$ from observations of \sneiip in the plateau phase alone; even having measurements of $v_\mathrm{Fe,50}$ does not break these degeneracies because it hardly contains any new information compared to $L_\mathrm{p,50}$. Conversely, this is the reason why one can use $v_\mathrm{Fe,50}$ to obtain the intrinsic bolometric plateau luminosity and thus potentially use \sneiip as standardised candles.

\begin{figure}
    \centering
    \includegraphics{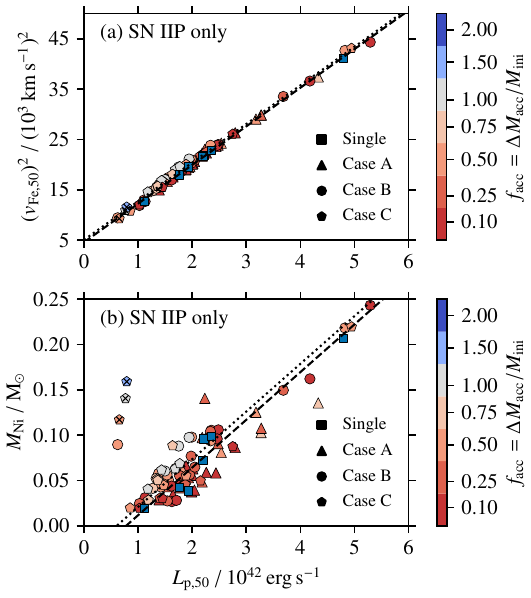}
    \caption{Ejected nickel mass $M_\mathrm{Ni}$ (a) and squared SN velocity $(v_\mathrm{Fe,50})^2$ (b) as a function of plateau luminosity $L_\mathrm{p,50}$ of our \sniip models. The different symbols are for single stars, and \casea, B and~C accretors with the accreted mass fraction $f_\mathrm{acc}$ indicated by the colours. Linear fits to the data of the single-star models (black dashed lines) are shown to guide the eye. Moreover, linear fits to the data of the accretor models with the same slopes as those for the single-star models are provided to indicate possible systematic differences (black dotted lines). Black crosses indicate models with likely significant LBV mass loss.}
    \label{fig:Lp-vej-MNi}
\end{figure}

Because $L_\mathrm{p,50}$ and $v_\mathrm{Fe,50}$ do not scale exactly as $L_\mathrm{p,50}\propto (v_\mathrm{Fe,50})^2$, we investigate the scatter in this relation in Fig.~\ref{fig:Lp-vej-MNi}a for our single-star models, and the \casea, B and~C accretors. The scatter around a one-to-one relation is minimal. However, there is a slight systematic offset between the single-star and the accretor models. A linear fit to the data of the single stars reveals a systematically higher $v_\mathrm{Fe,50}$ for the same $L_\mathrm{p,50}$ of the accretor models of $\Delta (v_\mathrm{Fe,50})^2 \approx 0.5\,(\kms)^2$; the offset increases to $\Delta (v_\mathrm{Fe,50})^2 \approx 0.9\,(\kms)^2$ when only considering accretors with the highest accretion fractions $f_\mathrm{acc}\geq0.75$. This offset is still small and only slightly broadens the $L_\mathrm{p,50}$--$v_\mathrm{Fe,50}$ relation such that accretors of binary mass transfer and stellar mergers are not expected to introduce large systematics in the observationally calibrated $L_\mathrm{p,50}$--$v_\mathrm{Fe,50}$ relation used for distance determination from \sneiip with the standardized candle method because of their different evolutionary histories.

Observationally, there is also a clear trend between the \sniip plateau luminosities and the nickel masses \citep[\eg][]{hamuy2003b, spiro2014a, pejcha2015a, gutierrez2017a, martinez2022b}. This relation is qualitatively in agreement with our model predictions (Fig.~\ref{fig:Lp-vej-MNi}b). We find a systematic offset between the single-star models and the \casea, B and~C accretors: the accretor models tend to have nickel masses larger by about $0.01\,\msun$ for the same $L_\mathrm{p,50}$ or, alternatively, smaller $L_\mathrm{p,50}$ for the same nickel masses. Furthermore, accretor models with larger $f_\mathrm{acc}$, \ie larger SN ejecta masses, at the same plateau luminosities generally have larger nickel masses (\cf Fig.~\ref{fig:Lp-vej-MNi}b at $L_\mathrm{p,50}=1\text{--}2\times10^{42}\,\mathrm{erg\,s^{-1}}$). This implies that the different evolutionary histories of Type IIP SN progenitors likely lead to a small albeit non-negligible broadening of the plateau luminosity--nickel mass relation. This trend can be understood as follows: for a given nickel mass, one has an explosion energy following the explosion energy--nickel mass relation found in our models (Eq.~\ref{eq:MNi-Eexpl-fit}) and generally in the literature (see Sect.~\ref{sec:nickel-proxy}). The explosion energy is one of the most important parameters setting the plateau luminosities of \sniip (\cf Eq.~\ref{eq:Lp}). In stars with a higher accretion fraction, the envelope and hence ejecta masses are larger such that these models then have systematically smaller plateau luminosities according to Eq.~(\ref{eq:Lp}).

In the above description, we have considered the total ejecta mass $M_\mathrm{ej}$ that includes the hydrogen- and helium-rich parts of the SN progenitor, following the fit formulae of \citet{goldberg2019a}. However, as shown by, \eg, \citet{dessart2019b} and \citet{fang2025a}, it is rather the hydrogen-rich ejecta mass that influences the light curves and spectral appearance of \sneiip strongest and not the total ejecta mass. For this reason, we also provide the hydrogen-rich envelope mass $M_\mathrm{H,env}$ defined as the total envelope mass above that location in a pre-SN star where the hydrogen mass fraction ${>}\,10^{-5}$ (see Tables~\ref{tab:sn-properties} and~\ref{tab:models}). The qualitative findings for the ejecta masses $M_\mathrm{ej}$ also apply to $M_\mathrm{H,env}$ but $M_\mathrm{ej}>M_\mathrm{H,env}$. Accretors of binary mass transfer and stellar mergers may have a higher helium abundance in their envelopes than genuine single stars. As the plateau phase of \sneiip is driven by hydrogen recombination, it may be advisable to consider the total hydrogen mass in the SN ejecta rather than $M_\mathrm{H,env}$ and $M_\mathrm{ej}$.

%
%
\section{\label{sec:comparison-observations}Comparison to observations}

In the following, we compare our models to observed \sneiip (Sect.~\ref{sec:SN-IIP}), \snesalike (Sect.~\ref{sec:long-risetime-sne}) and interacting and superluminous Type~II SNe (Sect.~\ref{sec:interacting-sne}). In this comparison, it is important to recall that our models are for a solar-like metallicity, while the observed SNe will be from progenitors of a range of metallicities. To first order, the metallicity mainly affects the stellar winds and thereby the final stellar masses. Hence, at lower (higher) metallicities, our models would, \eg, have larger (smaller) ejecta masses and lower (higher) $E_\mathrm{expl}/M_\mathrm{ej}$ ratios. 

\begin{table}
    \caption{\label{tab:sn-props-literature}Observationally inferred ranges of the explosion energies $E_\mathrm{expl}$, ejecta masses $M_\mathrm{ej}$, progenitor radii $R_\mathrm{cc}$ and nickel masses $M_\mathrm{Ni}$ of Type II SNe.}
    \centering
    \addtolength{\tabcolsep}{-2.0pt}
    \begin{tabular}{lcccc}
        \toprule
        Reference & $E_\mathrm{expl}$ & $M_\mathrm{ej}$ & $R_\mathrm{cc}$ & $M_\mathrm{Ni}$ \\
         & $(10^{51}\,\mathrm{erg\,s^{-1}})$ & $(\msun)$ & $(\rsun)$ & $(\msun)$ \\
        \midrule
        \multicolumn{5}{c}{\sneiip} \\
        \midrule
        H03 & $0.6\text{--}5.5$ & $14\text{--}56$ & $80\text{--}600$ & $0.016\text{--}0.260$ \\
        M17 & -- & -- & -- & $0.005\text{--}0.280$ \\
        M22 & $0.15\text{--}1.40$ & $8\text{--}14$ & $450\text{--}1000$ & $0.006\text{--}0.069$ \\
        UC24 & $0.25\text{--}10.20$ & $14\text{--}28$ & $35\text{--}1500$ & $0.006\text{--}0.110$ \\
        \midrule
        \multicolumn{5}{c}{\snesalike} \\
        \midrule
        T12\,\&\,16 & $0.6\text{--}11.3$ & $12\text{--}43$ & $30\text{--}350$ & $0.039\text{--}0.230$ \\
        P23 & $0.5\text{--}5.6$ & $15\text{--}40$ & $3\text{--}850$ & $0.040\text{--}0.230$ \\
        \bottomrule
    \end{tabular}
    \tablefoot{References are H03 \citep{hamuy2003b}, M17 \citep{muller2017a}, M22 \citep{martinez2022a}, UC24 \citep{utrobin2024a}, T12\,\&\,16 \citep{taddia2012a, taddia2016a}, \citep{pumo2023a}. When preparing the above data, the inferred properties of SN\,2009kf have been excluded from \citet{utrobin2024a}, and those of SN\,2004ek and OGLE073 from \cite{taddia2016a} and \citet{pumo2023a}. Among the \snesalike objects in T12\,\&\,16 and P23, there are large inferred stellar radii (${\gtrsim}\,1000\,\rsun$) for some long-rising SN light curves; excluding them leaves a sample of SNe with inferred progenitor radii $\lesssim 300\,\rsun$.}
\end{table}

\subsection{\label{sec:SN-IIP}Type IIP supernovae}

In the literature, one finds very different and to some extent contradictory ranges of inferred explosion energies, ejecta masses, progenitor radii and nickel masses of \sneiip (Table~\ref{tab:sn-props-literature}). The arguably most robust property is the nickel mass inferred from the tail of \sniip light curves \citep[see \eg][]{suwa2019a, anderson2019a}. The values of $0.005\text{--}0.280\,\msun$ reported by \citet{muller2017a} and based on a model of \citet{pejcha2015a} may serve as a benchmark. This range of nickel masses is in good agreement with our model predictions (Table~\ref{tab:sn-properties}). 

While $M_\mathrm{Ni}$ can be determined uniquely from the tail of the SN light curve, it is not possible to break the degeneracy between $E_\mathrm{expl}$, $M_\mathrm{ej}$ and $R_\mathrm{cc}$ from observations of the plateau phase of \sniip light curves and measurements of $v_\mathrm{Fe,50}$ \citep[][]{goldberg2019a, goldberg2020b}. The quite different ranges of these parameters inferred from \sniip observations may be partly because of implicit assumptions made to break the degeneracies. For example, \citet{martinez2022a} employ single-star models at solar metallicity to obtain an ejecta mass--progenitor radius relation from the stellar models in order to then break the degeneracy in $E_\mathrm{expl}$, $M_\mathrm{ej}$ and $R_\mathrm{cc}$. As \citet{martinez2022a} explain, the inferred values of $E_\mathrm{expl}$, $M_\mathrm{ej}$ and $R_\mathrm{cc}$ are then determined by their stellar models and have limited applicability.

To break the degeneracy, \citet{goldberg2019a} suggest constraining the radius of the SN progenitor, \eg, by characterising the SN progenitor star if it was observed or by modelling the SN light curve immediately after SN shock breakout (\ie during the first hours to days). Alternatively, one could try to constrain one of the degenerate parameters $E_\mathrm{expl}$, $M_\mathrm{ej}$ and $R_\mathrm{cc}$ by other means. In general, the explosion energies are linked to the nickel masses as explained in Sect.~\ref{sec:explosion-properties} (see also Fig.~\ref{fig:MNi-Eexpl}). Nickel masses can be determined observationally, \eg, from the tail of the SN light curve. Assuming there is a direct relation with the explosion energy, one could obtain $E_\mathrm{expl}$ independently and thereby break the degeneracy between $E_\mathrm{expl}$, $M_\mathrm{ej}$ and $R_\mathrm{cc}$. However, while we find a relatively tight relation between $E_\mathrm{expl}$ and $M_\mathrm{Ni}$ (Sect.~\ref{sec:nickel-proxy} and Fig.~\ref{fig:MNi-Eexpl}), the scatter is ${\approx}\, 0.3\times10^{51}\,\mathrm{erg}$ and might preclude an accurate enough inference of $E_\mathrm{expl}$ (Jared Goldberg, private communication).

While the \sniip explosion properties inferred by \citet{martinez2022a} cannot be compared to our accretor models for the reasons given above, we can compare them to those of our single-star models. The ranges of ejecta masses and progenitor radii agree well, but \citet{martinez2022a} find lower explosion energies and also lower nickel masses. Nickel masses and explosion energies are coupled, and the maximum nickel masses of \citet{martinez2022a} are much smaller than those reported by \citet{muller2017a}. This may suggest that the SN sample of \citet{martinez2022a} is biased towards low explosion energies\footnote{\citet{martinez2022a} also reconstruct an initial mass distribution of their SN progenitor stars that is significantly steeper than the typical \citet{salpeter1955a} stellar initial mass function. This could be a hint for missing physics in the applied SN progenitor models or a bias towards low initial masses in their SN sample.}. Our (single-stars) models do not reach down to the regimes of low iron core masses and electron-capture SNe, hence our minimum explosion energies and nickel masses are upper limits and larger than the minimum values found by \citet[][see also Tables~\ref{tab:sn-properties} and~\ref{tab:sn-props-literature}]{martinez2022a}. 

It is noteworthy that some observationally inferred ejecta masses of \sneiip (Table~\ref{tab:sn-props-literature}) exceed the maximum initial masses inferred for observed \sniip progenitors \citep[${\lesssim}\,20\,\msun$, see][]{smartt2009a, smartt2015a}. This apparent discrepancy may have (at least) two causes\footnote{The inference of high ejecta masses is likely also affected by the degeneracy issue described above \citep[\cf][]{goldberg2019a} and the general difficulty to infer ejecta masses while the \sniip light curves are more sensitive to the hydrogen-rich ejecta mass and not the total ejecta mass \citep[see \eg][]{dessart2019b}.}: (i) an observational bias that precluded the discovery of initially more massive \sniip progenitors \citep[\eg][]{beasor2024b} and (ii) binary mass accretors and stellar mergers. About half of all Type~II SNe are from stars that experienced mass accretion in a binary or merged with a companion \citep[\eg][]{podsiadlowski1992a, zapartas2019a}. Moreover, as explained in, \eg, \citet{farrell2020a}, \citet{temaj2024a} and \citetalias{schneider2024a}, the (mean) luminosities of \sniip progenitors are solely determined by the core masses of the progenitors and hold no information on the evolutionary history of the SN progenitor and thus the envelope and initial mass. Hence, binary products exploding with unusually large envelope masses and typical core masses may explain the inferred large ejecta masses of \sneiip and the low inferred initial masses of their progenitors. However, massive enough RSG SN progenitors are subject to large radial pulsations and thus luminosity variations \citep[see \eg][]{stothers1969a, wood1983a, heger1997a, yoon2010b, clayton2018a, bronner2025a}. These make comparisons to stellar models much more difficult and may preclude an accurate determination of the CO core masses of certain RSG SN progenitors.

Some \sneiip such as SN\,2015ba show unusually long plateau durations (${\approx}\,123\,\mathrm{d}$) and have rather high inferred ejecta masses \citep[${\approx}\,24\,\msun$;][]{dastidar2018a}. Such a high ejecta mass would signal a high initial mass of a single-star SN progenitor, implying a rather large oxygen core mass and thus oxygen-rich ejecta that should be visible in oxygen lines during the nebular phase. However, \citet{dastidar2018a} find insignificant levels of oxygen during the nebular phase in apparent contradiction with a massive SN progenitor. A past binary accretor (\eg stellar merger) may naturally explain this. The inferred nickel mass of SN\,2015ba of ${\approx}\,0.03\,\msun$ \citep{dastidar2018a} would indicate an SN progenitor with a central specific entropy of $s_\mathrm{c}/k_\mathrm{B} N_\mathrm{A} \approx 0.8\text{--}0.9$ and hence a rather small CO core mass of $M_\mathrm{CO}\lesssim 5\text{--}6\,\msun$ in our models (\cf Figs.~\ref{fig:Eexpl-MNi-vkick} and~\ref{fig:MCO-entropy}). The smaller CO core mass may explain the insignificant oxygen seen during the nebular phase and the past accretion/stellar merger could explain the large ejecta mass and hence long plateau duration. Similar situations are conceivable for other long-duration \sneiip.

The sub-group of low-luminosity \sneiip \citep[plateau luminosities lower by a factor of 5--10 compared to normal \sneiip, see \eg][]{pastorello2004a, spiro2014a, pumo2017a} with inferred low explosion energies (${\approx}\,10^{50}\,\mathrm{erg}$), small ejecta masses (${\approx}\,10\,\msun$ and possibly smaller), low nickel yields ($<0.01\,\msun$) and slow expansion velocities (factor of 2--3 slower than typical \sneiip) likely originate from the lower end of the initial mass range of core-collapse SN progenitors. Such stars and also electron-capture SN progenitors are not modelled here. Still, such explosions may occur in RSG and BSG progenitors in analogy to our more massive accretor models (\ie an electron-capture SN engine exploding a BSG with a high envelope mass). 

\citet{fang2025a} show that it is possible to infer the hydrogen-rich ejecta mass of \sneiip, and they find values in the range of ${\approx}\,2\text{--}11\,\msun$ for \sneiip. While the lower end agrees with our models, the high-mass end is smaller than what we predict (${\approx}\,26\,\msun$, Table~\ref{tab:sn-properties}). It is also smaller than what \citet{dastidar2018a} found for the long-plateau SN\,2015ba ($M_\mathrm{ej}\,{\approx}\,24\,\msun$), even when considering the difference between total and hydrogen-rich ejecta. A larger sample of \sneiip analysed with the methods outlined in \citet{fang2025a} may be needed to better understand the maximum hydrogen-rich ejecta mass.

\subsection{\label{sec:long-risetime-sne}SN1987A-like Type II SNe}

\snesalike events have long-rising, dome-shaped light curves and are sometimes also referred to as peculiar SN~II. As in \snesa \citep[see, \eg, reviews by][]{podsiadlowski1992c, podsiadlowski2017b}, their progenitors are likely compact BSGs and they occur at a rate of about 1--7\% of hydrogen-rich Type II SNe \citep{pastorello2012a, shivvers2017a}. The BSG progenitor of \snesa can neither be explained by a \casea nor a \caseb merger, while \snesalike events may be from such merged stars. To explain the BSG progenitor of \snesa, a \casec merger is required with sufficient mixing of helium out of the core into the envelope \citep{hillebrandt1989a, podsiadlowski1990a, podsiadlowski1992c, podsiadlowski2017a, podsiadlowski2017b, menon2017a}. In our accretor models, we do not apply such mixing which limits the range of accretion fractions $f_\mathrm{acc}$ from which we obtain BSG SN progenitors and affects the pre-SN position of our BSGs in the HR diagram \citepalias[for a detailed discussion of these points, we refer the reader to][]{schneider2024a}. 

In Table~\ref{tab:sn-props-literature}, we provide typical ranges of explosion energies, ejecta masses, nickel yields and SN progenitor radii inferred for \snesalikeshort SNe. Our choice of associating SN progenitors to \snesalike events by their position in the HR diagram naturally limits the SN progenitor radii to ${\lesssim}\,200\,\rsun$ (Table~\ref{tab:sn-properties} and Sect.~\ref{sec:hrd}). While this ad-hoc choice is arbitrary at first sight, it reflects the inferred radii of \snesalike progenitors quite well (${\lesssim}\,300\,\rsun$ with a few long-rising Type~II peculiar SNe having larger inferred progenitor radii of ${\gtrsim}\,1000\,\rsun$, \ie more typical of \sniip progenitors; see Table~\ref{tab:sn-props-literature}). This seems to generally support the idea that many \snesalike events are from BSG progenitors.

The observationally inferred nickel masses of ${\approx}\,0.04\text{--}0.23\,\msun$ (Table~\ref{tab:sn-props-literature}) appear to be consistent with the range of $0.02\text{--}0.24\,\msun$ predicted by our models (Table~\ref{tab:sn-properties}). There might be a slight tension for the lowest nickel masses where our models --- despite not covering the lowest iron core masses by design --- predict smaller values than what appears to have been observed\footnote{We note that the samples of \citet{taddia2012a}, \citet{taddia2016a} and \citet{pumo2023a} discussed here comprise $\mathcal{O}(10)$ \snesalikeshort SNe such that the quoted ranges of SN properties are subject to low number statistics.}. \citet{moriya2024a} show that exploding BSGs with $M_\mathrm{Ni}<0.01\,\msun$ do not give rise to long-rising, dome-shaped \snesalike light curves but are more similar to light curves of some intermediate-luminosity red transients. Hence, this may explain parts of this tension as an observational bias.

Because the nickel mass correlates with the explosion energy, our models predict lower minimum explosion energies ($0.2\times10^{51}\,\mathrm{erg}$) than what is inferred from observations ($0.5\times10^{51}\,\mathrm{erg}$). Moreover, some \snesalike events have large inferred explosion energies of $5\text{--}10\times10^{51}\,\mathrm{erg}$, which may be difficult to reach by standard neutrino-driven explosions \citep{janka2017a}. Additional energy sources such as another SN engine or circumstellar interaction may be required. Because the progenitors of \snesalikeshort SNe are likely stellar merger products, they may have highly magnetised cores at core collapse from the past coalescence \citep{schneider2019a, schneider2020a}. Neutrino-driven explosions from highly magnetised, non-rotating stars give rise to more energetic explosions \citep{matsumoto2022b, varma2023a}, which helps reduce this tension. At the extreme end of \snesalike SN~II, there is OGLE073 with an explosion energy of $12.4_{-5.9}^{+13.0}\times 10^{51}\,\mathrm{erg}$, an ejecta mass of $60_{-16}^{+42}\,\msun$ and a nickel mass of ${\geq}\,0.47\pm0.02\,\msun$ \citep{terreran2017a}. Such explosion energies point to a different explosion mechanism, and \citet{kozyreva2018a} argue that OGLE073 could be a pair-instability SN.

To obtain BSG SN progenitors, sufficiently high mass accretion is required such that the pre-SN stars have large envelope masses for their cores \citepalias{schneider2024a}. Hence, the SN ejecta masses of \snesalikeshort SNe are predicted to be higher than those of \sneiip. This does not seem to be confirmed by the compiled observational data (Table~\ref{tab:sn-props-literature}), but may be related to the difficulties of inferring robust explosion properties, similar to the situation in \sneiip as explained in Sect.~\ref{sec:SN-IIP}. As in the case of \sneiip, the ejecta masses of \snesalikeshort SNe appear to be generally more massive than the initial masses inferred for the progenitor stars of \sneiip \citep[${\lesssim}\,20\,\msun$, see][]{smartt2009a, smartt2015a}. This could be an indication of larger pre-SN masses in \snesalikeshort SNe compared to \sneiip, but, as noted above, the initial masses of \sniip progenitors are determined from single-star models and do not account for stellar mergers and accretors of binary mass transfer.

The rates of \snesalike SNe in comparison to iron core-collapse SNe hold important information on the frequency of BSG formation as a result of a past stellar merger. \citet{zapartas2019a} find that ${\approx}\,28\%$ of their binary stars exploding in hydrogen-rich Type~II SNe experience a merger of a post-MS star with an MS star. For a primordial binary star fraction of $f_\mathrm{B}=0.7$ \citep{sana2012a}, this implies that ${\approx}\,20\%$ of SN~II may be from such objects. If most of the observed \snesalikeshort SNe are from such mergers that additionally continued their post-merger evolution and final fate as BSGs, we can infer that a fraction of ${\approx}\,(1\text{--}7\%)/20\%{\approx}\,5\text{--}35\%$ of post-MS+MS mergers must result in long-lived BSGs given the observed \snesalikeshort SN rate of 1--7\% of Type~II SNe. We cannot compare this number quantitatively to our models because we cannot easily relate the accretion fractions to initial binary star configurations and the question of which initial binary star systems may experience post-MS+MS star mergers. It is, however, clear from our models that stars need to gain sufficient mass to become and explode as a BSG. Hence, a picture in which some 5--35\% of post-MS+MS star mergers become long-lived BSGs seems conceivable.

\subsection{\label{sec:interacting-sne}Interacting and superluminous SNe}

Interacting SNe are a diverse class of transients \citep[see \eg review by][]{smith2017a}. In these SNe, the fast-moving SN ejecta collides with the slower-moving circumstellar medium (CSM) previously ejected by the SN progenitor. The collision can decelerate the SN ejecta and thereby converts some SN kinetic energy into heat and radiation. Light emitted from the collision may ionise hydrogen in the CSM ahead of the shock which can give rise to the characteristic narrow hydrogen emission lines seen in many interacting SNe such as the prototypical \sneiin. Depending on the exact conditions, a considerable fraction of the kinetic energy can be converted into light, possibly giving rise to superluminous SNe \citep[SLSNe;][]{gal-yam2012a, howell2017a, moriya2018a}. The CSM interaction can mask the actual SN, and a wide range of SN explosion types may power interacting SNe (\eg core-collapse SNe, electron-capture SNe, (pulsational) pair-instability SNe and Type Ia SNe).

\sneiin make up about 4--10\% of hydrogen-rich SNe and about 3--7\% of core-collapse SNe \citep{li2011a, shivvers2017a}. Observationally, pre-SN mass loss rates of $10^{-4}\text{--}1\,\msun\,\mathrm{yr}^{-1}$ within the last $1\text{--}100\,\mathrm{yr}$ are inferred and the slower CSM typically moves at $100\text{--}1500\,\kms$ \citep{kiewe2012a}. Binary mass transfer, stellar mergers and common-envelope phases can eject the required amounts of mass and often lead to asymmetric, bipolar CSM structures. However, it is challenging to explain the timing of major mass ejection episodes from stellar mergers and common-envelope events and the SN explosion ${\lesssim}\,100\,\mathrm{yr}$ later unless the binary interaction itself causes the SN \citep[see \eg][]{hamuy2003a, chevalier2012a, schroder2020a}. Binary mass transfer just before a SN or still ongoing when the donor star explodes, may occur in a few per cent of all core-collapse SNe and is thus a promising channel for some \sneiin \citep[\eg][]{ercolino2024a}. Similarly, wave-driven mass loss induced by late nuclear burning might eject mass just prior to the final SN explosion and hence lead to \sniin characteristics \citep{arnett2011a, fuller2017a, wu2021a}; however, this mechanism seems to be more effective in hydrogen-poor SN progenitors \citep{wu2022b}. The \sniin characteristics are also met by exploding LBVs as further supported by the direct detection of very luminous ($L\sim10^6\,\lsun$) and yet compact, blue SN progenitors in SN~1961V, SN~2005gl, SN~2009ip and SN~2010jl \citep{gal-yam2007a, gal-yam2009a, smith2010a, kochanek2011a, smith2011a, foley2011a}. The latter, SN~2010jl, belongs to the class of SLSNe \citep{smith2017a}.

In standard single-star evolution, stars are not expected to end their lives as LBVs but to lose almost their entire hydrogen-rich envelope in strong mass loss and then end their lives as Wolf--Rayet stars, possibly giving rise to \sneibc. In this paper, we find that this is also true for binary mass accretors and stellar merger products that evolve as CSGs during most of core helium burning (\cf models marked by crosses in Fig.~\ref{fig:hrd-all}). Moreover, only very massive single stars (initially ${\gtrsim}\,30\text{--}40\,\msun$) are expected to become LBVs and only initially ${\gtrsim}50\,\msun$ may reach the very high SN progenitor luminosities inferred in SN~1961V, SN~2005gl, SN~2009ip and SN~2010jl\footnote{It is not clear whether the SN progenitors are caught during an LBV-like outburst during which they would appear overluminous or in a quiescent phase where the observed luminosity would be that of the star.}. Such massive stars are rather prone to collapsing into BHs and are thus not expected to explode in neutrino-driven SNe \citep[\eg][]{heger2003a, ertl2016a, schneider2021a, schneider2024a}. Hence, a different evolutionary path (\eg stellar mergers) and/or a different explosion mechanism (\eg pulsational PISNe) is required to explain interacting or (pulsational) pair-instability SNe from very luminous BSGs \citep[\cf][]{justham2014a, vigna-gomez2019a, renzo2020c, costa2022a, ballone2023a, schneider2024a}. Here, we consider the merger hypothesis within neutrino-driven SNe.

Some of our SN progenitor models labelled as \snesalike and \sneiin spend less than $10^4\,\mathrm{yr}$ inside the LBV regime of the HR diagram before exploding in a neutrino-driven SNe (Fig.~\ref{fig:hrd-all}). Their progenitor luminosities reach values of up to $\log\,L\,/\,\lsun\approx 6.1$ (Table~\ref{tab:sn-properties}) and are generally BSGs that have already burnt helium in their cores as such BSGs because of a past stellar merger. This agrees qualitatively well with the colours and luminosities inferred for the (possibly quiescent) progenitor stars of the observed (superluminous) \sneiin described above. As these models explode inside the LBV regime, they will likely have shed additional mass before core collapse and are thus excellent candidates to help understand some of the mentioned \sneiin. The ejecta masses of these models can be very high (up to ${\approx}\,80\,\msun$, Table~\ref{tab:sn-properties} and \citetalias{schneider2024a}), depending on how much mass the progenitors lose during their final evolution in the LBV region of the HR diagram. The explosion energies and nickel yields are rather comparable to those of our other models but also reach some of the highest values in our grid because these models are close to the compactness/entropy peak and high entropy signals a high explosion energy and nickel yield in our models (\cf Sects.~\ref{sec:explosion-properties} and~\ref{sec:sn-diversity}). In the rare situation that a large amount of mass is lost in an LBV giant eruption shortly before the SN (less than a few years), a large fraction of the SN kinetic energy may be converted into light, possibly giving rise to a Type~II SLSNe powered by CSM interaction.

The luminosities of BSG SN progenitors inside the LBV region of the HR diagram are given by their envelope mass \citepalias{schneider2024a}, and they have CO core masses of $7\text{--}9\,\msun$ (and up to $15\,\msun$ when neglecting strong LBV-like mass loss; Table~\ref{tab:sn-properties}). Hence, within the neutrino-driven SN mechanism, one may conclude that \sneiin from very bright SN progenitors are from stars with generally large CO core masses. This prediction may be testable in the future if oxygen lines in the nebular phase after the main SN has faded can be linked to their CO core mass, \eg, as is possible in stripped-envelope Type Ib/c and Type IIb SNe \citep[see \eg][]{jerkstrand2015a, fang2023a}.

Our models that likely experience strong LBV-like mass loss as a BSG or enhanced mass loss as a CSG for more than $10^4\,\mathrm{yr}$ may lose (almost) all of their hydrogen-rich envelopes and could then explode as stripped-envelope SNe of types IIL, IIb and Ib/c with possible signs of interaction (\eg SN~2006jc within the known class of SN~Ibn). According to the models, such SNe are from stars with CO core masses in the range ${\approx}\,7\text{--}15\,\msun$ and helium core masses of ${\approx}\,10\text{--}27\,\msun$. This implies large ejecta masses of up to ${\approx}\,25\msun$ in SNe with rather typical explosion energies of ${\approx}\,0.6\text{--}3.2\times10^{51}\,\mathrm{erg}$. If LBV mass loss was linked to instabilities in the hydrogen-rich envelope of stars, one would not expect such mass loss in helium stars and hence no SN~Icn via the formation channels discussed in this paper\footnote{There are formation channels for SN~Icn by binary mass loss from helium stars shortly before SNe \citep[see \eg][]{wei2024a}.}.

The rates of \sneiin and \snesalike are similar, and both are in the range of 1--10\% of hydrogen-rich SNe \citep[][]{li2011a, shivvers2017a}. Some of our post-MS+MS star mergers reach luminosities of ${\approx}\,10^6\,\lsun$ just before they explode inside the LBV region of the HR diagram and can thus explain the potentially very luminous SN progenitors of SN~1961V, SN~2005gl, SN~2009ip and SN~2010jl. Their formation rate must be significantly smaller than that of BSG SN progenitors of \snesalike SNe because of their higher masses. Hence, such merger products can only make up a small fraction of all \sneiin. While we cannot predict this contribution quantitatively, we can make educated guesses. If only a fraction $f_\mathrm{LBV-region}\,{\sim}\,0.1$ end up as a star in the LBV region prior to core collapse, one would infer a rate of $f_\mathrm{LBV-region}\times1\text{--}10\%\,{\approx}\,10^{-3}\text{--}10^{-2}$ of SN~II being \sneiin from a BSG SN progenitor. If another fraction of $f_\mathrm{giant-pre-SN-eruption}$ experiences a giant LBV eruption just before the SN, such events might only occur at a rate of $10^{-4}\text{--}10^{-3}$ of all core-collapse SNe (for $f_\mathrm{giant-pre-SN-eruption}=0.1$). Such low rates are typical for Type~II SLSNe from CSM interaction \citep[see \eg][]{moriya2018a}. Hence, BSGs from post-MS+MS star mergers could help understand these extreme events \citep[see also][]{justham2014a}. At a lower metallicity, some of our BSG progenitors avoiding long times in the LBV region of the HR diagram may have large enough cores to end in (pulsational) PISNe \citepalias[\cf][and see also \citealt{justham2014a, vigna-gomez2019a, renzo2020c, costa2022a, ballone2023a}]{schneider2024a}. Such objects might then further contribute to SLSNe.

As argued above, the majority of \sneiin are not related to LBV-like BSG progenitors. We now turn to exploding CSGs as contributors to \sneiin. Observationally, there is a mismatch between the luminosity of the most luminous \sniip progenitor \citep[$\log\, L/\lsun\,{\approx}\,5.1$;][]{smartt2009a, smartt2015a, rodriguez2022a} and the maximum luminosity of RSGs \citep[$\log\, L/\lsun\,{\approx}\,5.5$;][see also \citealt{temaj2024a}]{davies2018b, mcdonald2022a}, also known as the missing RSG problem\footnote{The significance of the missing RSG problem is constantly under debate as luminosity measurements of RSG progenitors are challenging \citep[\eg][]{beasor2024b} and the number of such detected progenitors is still rather small \citep[see, \eg,][]{davies2020a, rodriguez2022a}.}. Very luminous RSGs evolve into a luminosity-to-mass ratio regime where they may be subject to enhanced mass loss because of the proximity to the Eddington limit and envelope instabilities \citep[see \eg][]{clayton2018a}. In some stars, the mass loss may be so strong that the stars lose their entire hydrogen-rich envelopes and then explode in \sneibc. Others may still be in the process of losing their envelopes which could lead to \sneiin when an SN explodes into this enhanced wind. The rate of such SNe might actually be compatible with the observed \sniin rate. Let us assume that stars with initial masses beyond the compactness peak starting at $M_\mathrm{\xi-peak}\,{\approx}\,22\,\msun$ \citep{schneider2021a} and smaller than some upper limit $M_\mathrm{Ibc}=30\,\msun$ explode while enhanced CSG mass loss continuously erodes their envelopes. For an initial mass range of $\Delta M_\mathrm{\xi-peak}=2\,\msun$ of the compactness peak and a lower mass of $M_\mathrm{SN}=8\,\msun$ for the occurrence of core-collapse SNe, the expected fraction of \sneiin among Type~II SNe is 
\begin{align}
    r_\mathrm{IIn} &= \frac{M_\mathrm{Ibc}^{1-\gamma} - (M_\mathrm{\xi-peak} + \Delta M_\mathrm{\xi-peak})^{1-\gamma}}{  M_\mathrm{Ibc}^{1-\gamma} - (M_\mathrm{\xi-peak} + \Delta M_\mathrm{\xi-peak})^{1-\gamma} + M_\mathrm{\xi-peak}^{1-\gamma} - M_\mathrm{SN}^{1-\gamma}} \nonumber \\
    &\approx 7\%,
    \label{eq:sniin-rate}
\end{align}
where $\gamma=-2.35$ is the \citet{salpeter1955a} exponent of the stellar initial mass function. Such a rate would be compatible with the observed rate of \sneiin and could further help understand the missing RSG problem.

%
%
\section{\label{sec:conclusions}Conclusions}

We have studied the SN outcomes (\ie explosion energies, nickel yields, NS kick velocities and NS masses) and likely SN types of accretors of binary mass transfer and stellar mergers computed in \citetalias{schneider2024a}. The SN types \sniip, \snesalike and interacting \sniin were assigned to the models based on their position in the HR diagram. Our main results can be summarised as follows.
\begin{itemize}
    \item There is no obvious relation of the SN outcomes with the evolutionary history of our models (\ie single star, \casea, B and~C accretors). Instead, we find strong correlations of the SN outcomes with the central entropy of the stellar models at core-collapse. Similar correlations are found for other summary variables of the pre-SN core structure such as the compactness parameter or iron core mass. Hence, the exact evolutionary history of stars does not matter much for the SN explosion properties and outcomes as single stars and binary accretors can achieve similar pre-SN cores (albeit from different initial stellar masses). This does not imply that population averages of, \eg, SN explosion energies or nickel yields of single stars and binary accretors are also similar, but systematic differences can arise.
    \item We find that the nickel mass $M_\mathrm{Ni}$ is a proxy for the SN explosion energy $E_\mathrm{expl}$ and the NS remnant mass $M_\mathrm{NS,grav}$. Linear relations are found with root-mean-square deviations of ${\approx}\,0.3\times10^{51}\,\mathrm{erg}$ and ${\approx}\,0.05\,\msun$ for the $M_\mathrm{Ni}$--$E_\mathrm{expl}$ and $M_\mathrm{Ni}$--$M_\mathrm{NS,grav}$ relations, respectively. Quantitatively, the found relations depend on the stellar and SN models. Qualitatively, similar relations are also found in 3D simulations of neutrino-driven core-collapse SNe \citep[see, \eg,][]{burrows2024a}.
    \item Based on the assigned SN types, we summarise the properties of the SNe and their progenitors for single stars, \sneiip, \snesalike and \sneiin (Table~\ref{tab:sn-properties}) and, overall, find decent agreement with observations but also highlight some tensions (\eg, the minimum nickel mass for \snesalike in our models is lower than what is suggested by observations).
    \item In our models, \sneiip and \snesalike are from stars with similar CO core masses and pre-SN core structures, and therefore show a similar range of explosion properties. The main difference between these SN classes is the envelope structure in terms of mass and radius. \sneiip are from extended CSGs with convective envelopes, while the progenitors of \snesalike are compact BSGs with radiative envelopes. 
    \item Because of our assignment of \sneiin with more luminous SN progenitors compared to \sneiip and \snesalike, interacting \sneiin in our models are from stars with systematically larger CO core masses and hence different final central entropies and SN properties. For example, they generally have larger SN energies and higher Ni yields than what we find from \sneiip and \snesalike.
    \item Accretors of binary mass transfer and stellar mergers can lead to unusually long plateau durations of the SN light curves of \sneiip. At the same time, these models can have relatively small CO core masses and produce only small amounts of nickel but nevertheless eject large amounts of mass.
    \item Our models show a tight plateau luminosity--SN velocity relation, and accretors and mergers neither broaden nor induce offsets with respect to single-star models in this relation; such a relation is the basis for the standardised candle method \citep{hamuy2002a} for distance determinations using \sneiip. We further find a tight plateau luminosity--nickel mass relation in agreement with observations that is broadened by binary accretors and mergers; there is also a systematic offset with respect to our single-star models.
    \item Interacting \sneiin are a mixed bag of objects. Observationally, some extreme and superluminous SNe appear to be related to exploding ${\approx}\,10^6\,\lsun$ BSGs and we indeed find stellar merger models that explode in this location of the HR diagram and that may be subject to LBV-like mass loss just before the SN that could produce the circumstellar medium required for interacting SNe. 
    \item Given the upper luminosity threshold of observed RSGs and progenitors of \sneiip, these progenitors must have a high luminosity-to-mass ratio. It is thus conceivable that the most luminous CSG SN progenitors shed so much mass before the SN that they give rise to \sneiin. The predicted SN rate from such models could be compatible with the observed \sniin rate.
\end{itemize}
We have shown that accretors of binary mass transfer and stellar mergers contribute greatly to the observed diversity of hydrogen-rich \sneii. There remain many open ends, and the comparison of pre-SN stellar models to observations of SNe is difficult and challenged, \eg, by degeneracies such as in the case of \sniip light curves. Computing self-consistent SN light curves for stellar models may be a natural next step to further understand the large SN diversity.

%
%
\section*{\label{sec:data}Data availability}
The \mesa inlists, settings and some model output such as the pre-SN structures are available via Zenodo at \url{https://doi.org/10.5281/zenodo.10731998}, and the data from Table~\ref{tab:models} at \url{https://doi.org/10.5281/zenodo.15791044}.

\begin{acknowledgements}
We thank the anonymous reviewer for constructive feedback that helped improve this paper. FRNS and EL acknowledge support by the Klaus Tschira Foundation. This work has received funding from the European Research Council (ERC) under the European Union’s Horizon 2020 research and innovation programme (Grant agreement No.\ 945806) and is supported by the Deutsche Forschungsgemeinschaft (DFG, German Research Foundation) under Germany’s Excellence Strategy EXC 2181/1-390900948 (the Heidelberg STRUCTURES Excellence Cluster).
This research made use of NumPy \citep{oliphant2006a}, SciPy \citep{virtanen2020a}, Matplotlib \citep{hunter2007a} and Jupyter Notebooks \citep{kluyver2016a}.
\end{acknowledgements}

\bibliographystyle{aa}

\clearpage
\onecolumn
\appendix

\begin{landscape}
\section{\label{sec:model-data}Model data}

\tabcolsep=0.15cm
\begin{ThreePartTable}
    \renewcommand{\LTcapwidth}{\linewidth}
\begin{longtable}{ccccccccccccccccccc}
        \caption{\label{tab:models}Pre-SN and SN explosion properties of the presented stellar models.}\\
        \toprule
        $M_\mathrm{ini}$ & Case & $f_\mathrm{acc}$ & $t_\mathrm{cc}$ & $M_\mathrm{cc}$ & $M_\mathrm{CO}$ & $s_\mathrm{c}$ & SN & $M_\mathrm{rm}$ & $M_\mathrm{ej}$ & $M_\mathrm{H,env}$ & Fall- & $\log\,L_\mathrm{cc}$ & $\log\,T_\mathrm{eff,cc}$ & $R_\mathrm{cc}$ & $\delta t_\mathrm{LBV}$ & $E_\mathrm{expl}$ & $M_\mathrm{Ni}$ & $v_\mathrm{kick}$ \\
        ($\msun$) &  &  & (Myr) & ($\msun$) & ($\msun$) & ($N_\mathrm{A} k_\mathrm{B}$) & Type & ($\msun$) & ($\msun$) & ($\msun$) & back? & ($\log\,\lsun$) & ($\log\,\mathrm{K}$) & ($\rsun$) & $(10^4\,\mathrm{yr})$ & ($10^{51}\,\mathrm{erg}$) & ($\msun$) & ($\kms$) \\
        \midrule
        \endfirsthead
        \caption{continued.}\\
        \toprule
        $M_\mathrm{ini}$ & Case & $f_\mathrm{acc}$ & $t_\mathrm{cc}$ & $M_\mathrm{cc}$ & $M_\mathrm{CO}$ & $s_\mathrm{c}$ & SN & $M_\mathrm{rm}$ & $M_\mathrm{ej}$ & $M_\mathrm{H,env}$ & Fall- & $\log\,L_\mathrm{cc}$ & $\log\,T_\mathrm{eff,cc}$ & $R_\mathrm{cc}$ & $\delta t_\mathrm{LBV}$ & $E_\mathrm{expl}$ & $M_\mathrm{Ni}$ & $v_\mathrm{kick}$ \\
        ($\msun$) &  &  & (Myr) & ($\msun$) & ($\msun$) & ($N_\mathrm{A} k_\mathrm{B}$) & Type & ($\msun$) & ($\msun$) & ($\msun$) & back? & ($\log\,\lsun$) & ($\log\,\mathrm{K}$) & ($\rsun$) & $(10^4\,\mathrm{yr})$ & ($10^{51}\,\mathrm{erg}$) & ($\msun$) & ($\kms$) \\
        \midrule
        \endhead
        \bottomrule
        \endfoot
         11.0 & Single & -- & 22.4 & 10.0 & 2.0 & 0.78 & II & 1.30 & 8.6 & 6.8 & 0 & 4.76 & 3.52 & 718 & 0.0 & 0.45 & 0.02 & 216 \\
\midrule
         11.0 & \caseae & 0.10 & 21.6 & 11.0 & 2.1 & 0.81 & II & 1.33 & 9.6 & 7.6 & 0 & 4.80 & 3.52 & 757 & 0.0 & 0.98 & 0.04 & 507 \\
         11.0 & \caseae & 0.25 & 21.6 & 12.0 & 2.8 & 0.86 & II & 1.42 & 10.4 & 7.9 & 0 & 4.95 & 3.52 & 912 & 0.0 & 0.96 & 0.05 & 479 \\
         11.0 & \caseae & 0.50 & 21.5 & 13.0 & 3.9 & 0.90 & II & 1.45 & 11.3 & 7.3 & 0 & 5.12 & 3.53 & 1059 & 0.0 & 0.70 & 0.05 & 343 \\
         11.0 & \caseae & 0.75 & 21.0 & 13.3 & 5.3 & 0.91 & II & 1.46 & 11.6 & 6.2 & 0 & 5.28 & 3.57 & 1076 & 0.0 & 2.18 & 0.14 & 1426 \\
         11.0 & \caseae & 1.00 & 20.6 & 13.1 & 6.6 & 1.14 & BH & 13.1 & 0.0 & 4.6 & 0 & 5.40 & 3.61 & 1013 & 0.0 & 0.00 & 0.00 & 0 \\
         11.0 & \caseae & 1.50 & 20.1 & 14.1 & 9.3 & 0.95 & II & 1.59 & 12.3 & 2.6 & 0 & 5.62 & 3.73 & 758 & 1.4 & 1.07 & 0.13 & 631 \\
         11.0 & \caseae & 2.00 & 19.6 & 16.0 & 12.4 & 1.05 & II & 1.73 & 14.0 & 1.2 & 0 & 5.76 & 4.12 & 148 & 48.7 & 1.72 & 0.17 & 993 \\
\end{longtable}
    \end{longtable}
    \vspace{-0.8cm}
    \begin{minipage}[t]{0.95\linewidth}
        \tablefoot{The provided data are the initial mass $M_\mathrm{ini}$, the model case (single or accretor including an indication of the evolutionary stage at accretion), the fraction $f_\mathrm{acc}$ of accreted mass in units of $M_\mathrm{ini}$, the lifetime $t_\mathrm{cc}$ until core collapse, the final stellar mass $M_\mathrm{cc}$, the CO core mass $M_\mathrm{CO}$ at core collapse, the central entropy $s_\mathrm{c}$ at core collapse, the SN type, the gravitational compact-object remnant mass $M_\mathrm{rm}$, the SN ejecta mass $M_\mathrm{ej}$, the mass of the hydrogen-rich envelope $M_\mathrm{H,env}$ of the SN progenitor, a flag indicating the likely occurrence of SN fallback, the luminosity $\log\,L_\mathrm{cc}$, effective temperature $T_\mathrm{eff,cc}$ and radius $R_\mathrm{*}$ at core collapse, the time $\delta t_\mathrm{LBV}$ spent in the LBV-region of the HR diagram, the SN explosion energy $E_\mathrm{expl}$, the ejected nickel mass $M_\mathrm{Ni}$ and the SN kick velocity of a formed NS. The full table is available in electronic form via Zenodo at \url{https://doi.org/10.5281/zenodo.15791044}.}
    \end{minipage}
\end{ThreePartTable}
\end{landscape}

\end{document}